\documentclass[aps,prb,reprint,amsmath,amssymb]{revtex4-1}
\usepackage{dcolumn}% Align table columns on decimal point
\usepackage{bm}% bold math
\usepackage[utf8]{inputenc}

\usepackage{amsfonts}

\usepackage{ifpdf}
\usepackage{bm}
\usepackage{latexsym}
\usepackage{color}
\usepackage{pstricks}
\usepackage{array}

\usepackage{ifthen}
\usepackage{ifpdf}

\usepackage{latexsym}
\usepackage{amsmath}
\usepackage{amssymb}
\usepackage{bm}
\numberwithin{equation}{section}

\ifpdf
\usepackage{graphicx}
\usepackage{epstopdf}
\usepackage{float}
\else
\usepackage{graphicx}
\usepackage{epsfig}
\usepackage{float}
\fi
\usepackage{subfigure} %use this instead of caption of subcaption packages?

\newcommand{\LHx}{{{\rm {\text{LiHo}_{\text{x}}\text{Y}_{\text{1-x}}\text{F}_{\text{4}}}  }}}
\newcommand{\LH}{{{\rm {\text{LiHoF}_{\text{4}}}     }}}

\begin{document}
\title{Direct Measurement of Random Fields in the $\LHx$ Crystal}

\author{Yoav G. Pollack}
\author{Moshe Schechter}
\affiliation{Department of Physics, Ben-Gurion University of the Negev, Beer Sheva 84105, Israel}

\date{\today}
%%%%%%%%%%%%%%%%%%%%%%%%%%%%%%%%%%%%%%%%%%%%%%%%%%%%%%%%%%%%%%%
\begin{abstract}
The random field Ising model (RFIM) is central to the study of disordered systems. Yet, for a long time it eluded realization in ferromagnetic systems because of the difficulty to produce locally random magnetic fields. Recently it was shown that in anisotropic dipolar magnetic insulators, the archetypal of which is the $\LHx$ system, the RFIM can be realized in both ferromagnetic and spin glass phases. The interplay between an applied transverse field and the offdiagonal terms of the dipolar interaction produce effective longitudinal fields, which are random in sign and magnitude as a result of spatial dilution.
In this paper we use exact numerical diagonalization of the full Hamiltonian of Ho pairs in $\LHx$ to calculate the effective longitudinal field beyond the perturbative regime. In particular, we find that nearby spins can experience an effective field larger than the intrinsic dipolar broadening (of quantum states in zero field) which can therefore be evidenced in experiments. We then calculate the magnetization and susceptibility under several experimental protocols, and show how these protocols can produce direct measurement of the effective longitudinal field.
\end{abstract}\maketitle
%%%%%%%%%%%%%%%%%%%%%%%%%%%%%%%%%%%%%%%%%%%%%%%%%%%%%%%%%%%%%%%

%\pacs{75.50.Lk,75.10Nr,75.50Dd,75.10Jm}

\section{Introduction}

The random field Ising model (RFIM) has been central to the research of disordered systems ever since the seminal work of Imry and Ma\cite{ImryMa} who have shown that below a lower critical dimension of two the ferromagnetic (FM) phase is unstable to an infinitesimal random field. Whereas the RFIM provides a simple and adequate description for a plethora of problems across scientific disciplines, its realization in FM systems was hindered for a long time  because of the difficulty to produce magnetic fields which are random on short length scales. Experimentally, effects of the random field on e.g. the FM-paramagnetic (PM) phase transition were thoroughly studied using dilute antiferromagnets (DAFM) in a constant field (see Ref.~\onlinecite{Belanger:98} and refs. therein), as these systems were shown\cite{FishmanAharony} to be equivalent to the RFIM near criticality.

Recently it was shown that dilute anisotropic dipolar insulators in an applied magnetic field transverse to the easy (Ising) axis constitute a realization of the RFIM in both their FM and spin glass phases\cite{SchechterLaflorencie,TabeiGingras,Schechter}. In such systems, the archetypal of which is $\LHx$, the interplay between the applied transverse field and the offdiagonal elements of the dipolar interaction transforms spatial disorder into an effective random field in the longitudinal (Ising axis) direction. The offdiagonal dipolar terms cannot be neglected despite the strong Ising anisotropy because they break the $Z_2$ symmetry of $S_z \rightarrow -S_z$. Note, that in the absence of an applied magnetic field, $Z_2$ symmetry is protected by time reversal, i.e. $S \rightarrow -S$, but once the $Z_2$ symmetry is no longer protected by time reversal, random fields become generic as in non-magnetic systems\cite{Schechter:09}.

For the pure $\LH$ system, the effective longitudinal field at each spin site is zero, because contributions from all other spins cancel exactly. However, upon dilution cancelation is not exact, and the net longitudinal field at each populated site depends on the (random) position of the other spins.
We stress here that the phrase "effective longitudinal field" is reserved in this paper for the term that appears explicitly in the effective low energy Hamiltonian (third term in Eq. \ref{eq:fullIsing}), breaking time reversal symmetry. This is to be differed from the mean field effective fields exerted by the random interaction itself (first term in Eq. \ref{eq:fullIsing}).
The typical magnitude of the effective longitudinal fields depends on the concentration $x$; it is linear in $x$ for $x \ll 1$ \cite{SchechterLaflorencie} and proportional to $1-x$ for $1-x \ll 1$,\cite{Schechter} deep in the FM phase.

The possibility to study the RFIM in FM systems has advantages in comparison to its study in the DAFM, e.g. (i) a uniform tunable parallel magnetic field can be applied, which allows susceptibility measurements, and study of phenomena such as Barkhausen noise\cite{SDM01,CD03}. This is in contrast to the DAFM, where an effective field parallel to the staggered magnetization can not be applied. (ii) The RFIM can be studied away from criticality (DAFM in a field are equivalent to the RFIM only near criticality).

Since the theoretical prediction of its realization in anisotropic dipolar magnets, the RFIM was studied experimentally in both the SG and FM phases of the $\LHx$ system\cite{SilevitchNature, Barbara2007,SilevitchPRL,Silevitch16022010}. In particular, a peculiar dependence of the critical temperature on the random field was observed at $x=0.44$. This behavior was recently shown to be a result of the proximity to the SG phase, and the novel disordering it induces in the presence of a random field\cite{ATKS:12}.

In this paper we consider $\LHx$ in the extremely dilute regime, $x \ll 1$. In this regime, except at ultra low temperatures, physics is dominated by single spins, and by rare nearby spin pairs which have intra-pair interaction far larger than the typical spin-spin interactions in the system. Contributions of single spin tunneling and pair cotunneling to the hysteresis magnetization were reported in Refs.\cite{Giraud2001,Giraud2003}. Considering such nearby pairs, we calculate, using exact diagonliazation of the full two-Ho system, the effective longitudinal field for each pair given the relative position of the two spins. Our calculations extend the perturbative results in Ref.~\onlinecite{SchechterLaflorencie} to large fields. In addition, we calculate the contribution to the magnetization and susceptibility of various Ho pairs, as function of an applied field in both the longitudinal and transverse direction. Based on these calculations we describe experimental protocols that allow the measurement of the effective longitudinal field for given pairs, as these effective fields are manifested in shifted susceptibility resonances. Such experiments will constitute a direct measurement of the microscopic random field, rather than its macroscopic consequences.
\par
The structure of the paper is as follows: In Sec. \ref{LHx} we discuss the properties of the $\LHx$ crystal. In Sec. \ref{Random Effective Fields} we present the numerical technique and the calculation of the effective random fields. In Sec. \ref{Measurement} we present the calculations of the magnetization and susceptibility and propose experimental protocols for measuring the random fields. Our results are summarized in Sec.  \ref{conclusions}. App. \ref{Perturbative Expansion} reviews the analytic perturbative derivation of the random field and compares it to our numerical results. In App. \ref{numerics} we describe the numerical calculation in some detail.

%%%%%%%%%%%%%%%%%%%%%%%%%%%%%%%%%%%%%%%%%%%%%%%%%%%%%%%%%%%%%%%%%%%%%%%%%%%%%
%Figure: LiHoF4 unit cell
\begin{figure}[h]
  \centering
  \includegraphics[width=0.3\textwidth]{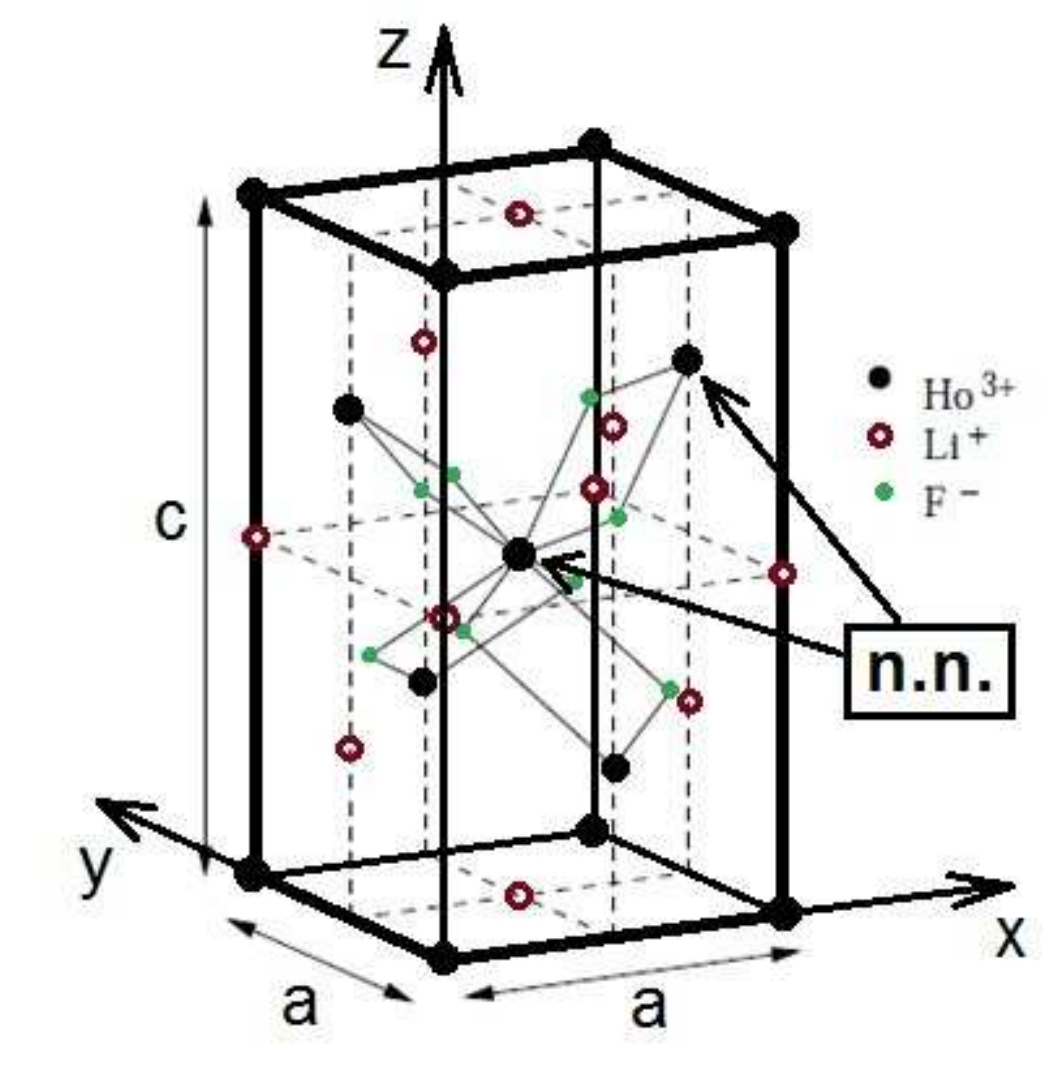}
  \caption{The unit cell for $\LH$. $a=5.175$\AA , $c=10.75$\AA. n.n. indicates Nearest lattice Neighbors. Illustration adapted from Gingras and Henelius \cite{acCellSize}. }
  \label{fig:LihoF4_cell}
\end{figure}
%%%%%%%%%%%%%%%%%%%%%%%%%%%%%%%%%%%%%%%%%%%%%%%%%%%%%%%%%%%%%%%%%%%%%%%%%%%%%

\section{$\LHx$}
\label{LHx}
\par
The $\LH$ crystal (see Fig.\ref{fig:LihoF4_cell}) is a realization of a dipolar Ising magnet. Disorder is introduced through dilution of the highly magnetic Ho sites (${\mu_{Ho}=10.6 \mu_B}$)\cite{Coqblin} by the practically non-magnetic Yttrium ions (${\mu_{Y}=0.14 \mu_B}$) to produce $\LHx$ with Ho concentration $x$.
The "free" trivalent Ho ion has the configuration ${^5 I_8}$. The Crystal Field (CF) generated by the electrostatic potential of the crystal, given by\cite{Chakraborty}
%%%%%%%%%%%%%%%%%%%%%%%%%%%%%%%%%%%%%%%%%%%%%%%%%%%%%%%%%%%%%%%%%%%%%%%%%%%%%%%%%%%%%
\begin{equation}
\begin{aligned}
H_{CF}=&B_2^0O_2^0+B_4^0O_4^0+B_6^0O_6^0+B_4^4(C)O_4^4(C)\\
&+B_6^4(C)O_6^4(C)+B_6^4(S)O_6^4(S)
\end{aligned}
\label{eq:CF}
\end{equation}
%%%%%%%%%%%%%%%%%%%%%%%%%%%%%%%%%%%%%%%%%%%%%%%%%%%%%%%%%%%%%%%%%%%%%%%%%%%%%%%%%%%%%
partially breaks the 17-fold degeneracy (${J^z=-8,-7,...,7,8}$) and translates to a large uniaxial magnetic anisotropy along the z axis.

Here the $O_n^l$ are Stevens' operator equivalents \cite{Stevens} and the $B_n^m$ coefficients are crystal field parameters obtained through fitting to spectroscopic and neutron scattering experiments\cite{Ronnow}.
The $O_4^4(C)$, $O_6^4(C)$ and $O_6^4(S)$ terms break the easy z axis symmetry and couple free-ion states with ${\Delta J^z=\pm4}$. This produces a doubly degenerate ground Ising state ${|\uparrow>}$ and ${|\downarrow>}$ with ${<J^z>=\pm5.5}$. The first excited state ($\Gamma_2$) is well above the ground states at $\Omega_0=10.8$K .

%HyperFine
\par
The Ho ion has a nuclear spin of $7/2$, and an untypically large hyperfine (HF) interaction between the electronic and nuclear angular moments. The HF interaction can be conveniently separated into two parts:
%%%%%%%%%%%%%%%%%%%%%%%%%%%%%%%%%%%%%%%%%%%%%%%%%%%%%%%%%%%%%%%%%%%%%%%%%%%%%
%Equation: HyperFine
\begin{equation}
\begin{aligned}
H_{HF}&=A_J\sum_i\vec{I}_i\cdot\vec{J}_i \\
&=A_J\sum_i I_i^z \cdot J_i^z+\frac{A_J}{2}\sum_i(I_i^ +\cdot J_i^- +I_i^- \cdot J_i^+)\\
&=H_{HF}^\parallel+H_{HF}^\perp
\end{aligned}
\label{eq:HF}
\end{equation}
%%%%%%%%%%%%%%%%%%%%%%%%%%%%%%%%%%%%%%%%%%%%%%%%%%%%%%%%%%%%%%%%%%%%%%%%%%%%%
with ${I=\frac{7}{2}}$ and $A_J=0.039$K \cite{Giraud2003}.
\par
\begin{figure}[h]
  \centering
  \includegraphics[width=0.5\textwidth]{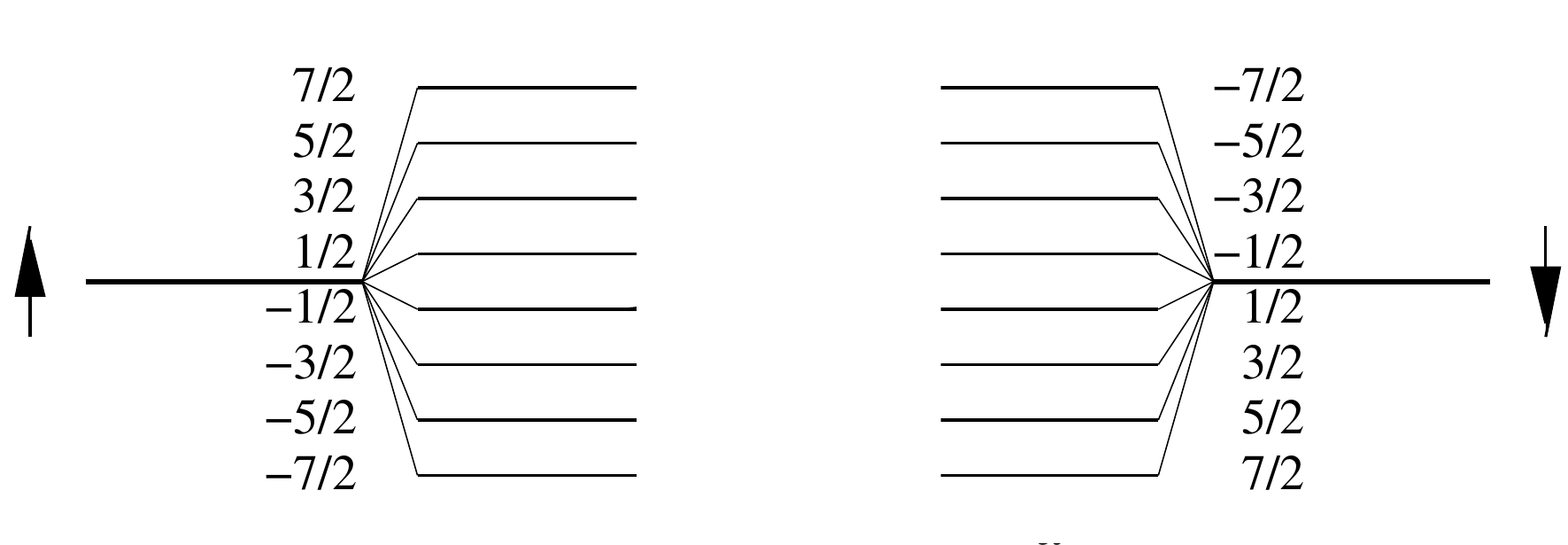}
  \caption{Splitting of the electronic Ising states into electro-nuclear states due to the "longitudinal" part of the HF interaction ($H_{HF}^{\parallel}$). Illustration taken from Schechter and Stamp \cite{SchechterStamp2008} }
  \label{fig:HF}
\end{figure}

The "longitudinal" part of the interaction ${H_{HF}^{\parallel}}$ splits each of the electronic ground doublet into 8 equidistant electro-nuclear levels  ${\Delta E\simeq 215}$mK each with its own $I^z$ (Fig. \ref{fig:HF}). These electro-nuclear states can be grouped in degenerate time reversal pairs, i.e. $\mid \uparrow,I_z \rangle$; $\mid \downarrow -I_z \rangle$, of which the lowest pair can be treated as the new Ising states.
%%%%%%%%%%%%%%%%%%%%%%%%%%%%%%%%%%%%%%%%%%%%%%%%%%%%%%%%%%%%%%%%%%%%%%%%%%%%%
%Figure: HF structure
%%%%%%%%%%%%%%%%%%%%%%%%%%%%%%%%%%%%%%%%%%%%%%%%%%%%%%%%%%%%%%%%%%%%%%%%%%%%%
The "transverse" part of the HF interaction combined with a transverse magnetic field under $2$T can only couple these time reversal pair states weakly (see Fig. 4 in Ref. \onlinecite{SchechterStamp2008}).

%longitudinal field
An applied longitudinal magnetic field, splits the degeneracy between the electro-nuclear doublets linearly (Land\'e g factor ${g_L=\frac{5}{4}}$).
%%%%%%%%%%%%%%%%%%%%%%%%%%%%%%%%%%%%%%%%%%%%%%%%%%%%%%%%%%%%%%%%%%%%%%%%%%%%%
%Equation: Energy gap with longitudinal field
\begin{equation}
\begin{aligned}
\Delta E&= g_L \mu_B B^z <J^z_{\uparrow}>-g_L \mu_B B^z <J^z_{\downarrow}>\\
&=2g_L \mu_B B^z <J^z_{\uparrow}>
\label{eq:longi_deltaE}
\end{aligned}
\end{equation}
%%%%%%%%%%%%%%%%%%%%%%%%%%%%%%%%%%%%%%%%%%%%%%%%%%%%%%%%%%%%%%%%%%%%%%%%%%%%%

The inter-ionic interaction between two Ho ions is composed of both superexchange (AFM) and dipolar components. The superexchange interaction is given by
%%%%%%%%%%%%%%%%%%%%%%%%%%%%%%%%%%%%%%%%%%%%%%%%%%%%%%%%%%%%%%%%%%%%%%%%%%%%%
\begin{equation}
H_{superexchange}= \mathop{ \sum_{ i \neq j}}  U(\vec{r}_{ij}) \vec{J_i} \vec{J_j} \, ,
\label{eq:superexchange}
\end{equation}
%%%%%%%%%%%%%%%%%%%%%%%%%%%%%%%%%%%%%%%%%%%%%%%%%%%%%%%%%%%%%%%%%%%%%%%%%%%%%
where the value of ${U(\vec{r_{ij}})}$ was found for the nearest lattice neighbor (n.n.) pairs ($U_{n.n.}=5.6$mK) and ${\text{2}^{\text{nd}}}$ n.n. pairs ($U_{2nd}=0.33$mK) through specific heat measurements \cite{superexchange}. The dipolar interaction is given by
%%%%%%%%%%%%%%%%%%%%%%%%%%%%%%%%%%%%%%%%%%%%%%%%%%%%%%%%%%%%%%%%%%%%%%%%%%%%%
\begin{equation}
\begin{aligned}
H_{dipolar}&= \mathop{ \sum_{ i \neq j}}_{ \alpha, \beta = x,y,z}  V_{ij}^{\alpha \beta} J_i^{\alpha} J_j^{\beta}& \\
= \sum_{ \alpha, \beta = x,y,z} \sum_{ j \neq i} \sum_i& \frac{1}{2} g_L^2 \mu_B^2 \frac{\mu_0}{4 \pi} \frac{|r_{ij}|^2 \delta_{\alpha , \beta}-3r_{ij}^{\alpha} r_{ij}^{\beta}}{|r_{ij}|^5} &J_i^{\alpha} J_j^{\beta} \,
\end{aligned}
\label{eq:dipolar}
\end{equation}
%%%%%%%%%%%%%%%%%%%%%%%%%%%%%%%%%%%%%%%%%%%%%%%%%%%%%%%%%%%%%%%%%%%%%%%%%%%%%
and is FM or AFM depending on the spatial alignment of the two interacting ions.
The superexchange interaction is small in comparison to the dipolar interaction even for n.n. pairs\cite{Cooke,Chakraborty,acCellSize,Reich}, and lacks off-diagonal (e.g. xz) terms. Thus, the effect of the superexchange interaction on the effective longitudinal fields is small, and is neglected hereafter.

Keeping only the dominating zz term of dipolar interactions (all other terms of the dipolar interaction involve excited single Ho electronic states, and are thus effectively reduced), the $\LHx$ system becomes a realization of the Ising model. The addition of an applied transverse magnetic field $B^x$, which allows for quantum fluctuations between the low energy single Ho Ising states, renders an effective transverse field. Furthermore, upon dilution, the combined effect of the transverse field and the offdiagonal terms of the dipolar interactions results in an effective random field\cite{SchechterLaflorencie,Schechter} (see App. \ref{Perturbative Expansion}. for details and a definition of the constant $\eta$).

%%%%%%%%%%%%%%%%%%%%%%%%%%%%%%%%%%%%%%%%%%%%%%%%%%%%%%%%%%%%%%%%%%%%%%%%%%%%%
%Equation: Perturbation B^z_eff leading term
\begin{equation}
 B^z_{k,eff}=\frac{4 \eta} {\Omega_0} \mathop{ \sum_{ i \neq k}} V_{ki}^{zx}  B^x \, .
\label{RFSL}
\end{equation}
%%%%%%%%%%%%%%%%%%%%%%%%%%%%%%%%%%%%%%%%%%%%%%%%%%%%%%%%%%%%%%%%%%%%%%%%%%%%%

Thus one can study in the $\LHx$ the RFIM in the presence of an effective transverse field $\Delta$ and a constant longitudinal field $B^z$, all effective fields being tunable by the choice of the applied longitudinal and transverse field, and Ho concentration\cite{Schechter}

%%%%%%%%%%%%%%%%%%%%%%%%%%%%%%%%%%%%%%%%%%%%%%%%%%%%%%%%%%%%%%%%%%%%%%%%%%%%%
%Equation: Effective Hamiltonian
\begin{equation}
H=\sum_{ij} V_{ij}^{eff} \tau_i^z \tau_j^z + \Delta \sum_i \tau_i^x + \gamma \sum_i  B^z_{i,eff} \tau_i^z +\gamma B^z \sum_i \tau_i^z .
 \label{eq:fullIsing}
\end{equation}
%%%%%%%%%%%%%%%%%%%%%%%%%%%%%%%%%%%%%%%%%%%%%%%%%%%%%%%%%%%%%%%%%%%%%%%%%%%%%
Here $\tau$ represents an effective spin half, $\gamma=2 g_L \mu_ B\langle J^z \rangle$ and $B^z_{i,eff}$ is given by Eq.(\ref{RFSL}).

For small applied transverse fields the effective random field dominates over the effective transverse field $\Delta$, as the former is linear in $B^x$ whereas the latter is higher order in the field\cite{Schechter}. Furthermore, for low energies, where the relevant Ising states are the electro-nuclear states $\mid \uparrow,I_z \rangle$; $\mid \downarrow -I_z \rangle$, the effective transverse field is significantly reduced as a consequence of the need to also flip the nuclear state\cite{SchechterStamp2005,SchechterStamp2008}. Thus for small applied transverse fields the Hamiltonian in Eq.(\ref{eq:fullIsing}) practically reduces to the classical RFIM. This holds for any Ho concentration $x < 1$, and for $x>0.3$ it yields a FM RFIM.

\section{Random Effective Fields for Ho pairs}
\label{Random Effective Fields}
The random field has profound effects on the system, in both its spin glass and ferromagnetic phases\cite{SchechterLaflorencie,TabeiGingras,SSL:07,Schechter,Wu1993,SilevitchNature,SilevitchPRL}. However, its direct measurement requires special consideration. This is because the energy change when flipping spins is a sum of the random fields exerted on the flipped spins, and their (random) interactions with the rest of the system. However, in the very dilute regime, the situation is favorable, since pairs of spins separated by a distance significantly smaller than the typical distance between spins interact predominantly within the pair, and only weakly with the rest of the system. We now analyze the effective longitudinal fields of such pairs of spins.

Let us consider a pair of Ho ions with a relative location dictating a FM dipolar interaction. Its two degenerate ground states are then $\mid \uparrow -\frac{7}{2} \uparrow -\frac{7}{2} \rangle$ and $\mid \downarrow \frac{7}{2} \downarrow \frac{7}{2}\rangle$. Upon the application of a transverse field, the effective (random) longitudinal field given in Eq.(\ref{RFSL}) for any number of spins reduces to the field exerted by each spin on the other. This longitudinal field is identical for both spins, and its magnitude and sign depends on the relative location of the two spins. It is directly related to the energy splitting of the degeneracy by the transverse field:

%%%%%%%%%%%%%%%%%%%%%%%%%%%%%%%%%%%%%%%%%%%%%%%%%%%%%%%%%%%%%%%%%%%%%%%%%%%%%
%Equation: Effective Field
\begin{equation}
B^z_{eff}(B^x) \equiv \frac{\left[ E(\mid \uparrow -\frac{7}{2} \uparrow -\frac{7}{2}\rangle) - E(\mid \downarrow \frac{7}{2}\downarrow \frac{7}{2}\rangle)\right]}{2 g_L \mu_B (<J^z_{\uparrow}>-<J^z_{\downarrow}>)} \, .
\label{eq:effective_field}
\end{equation}
%%%%%%%%%%%%%%%%%%%%%%%%%%%%%%%%%%%%%%%%%%%%%%%%%%%%%%%%%%%%%%%%%%%%%%%%%%%%%

For small $B^x$, the calculation of ${ E(\mid \uparrow -\frac{7}{2} \uparrow -\frac{7}{2}\rangle)} - {E(\mid \downarrow \frac{7}{2}\downarrow \frac{7}{2}\rangle)}$ can be calculated perturbatively (Eq.(\ref{RFSL}), App.\ref{Perturbative Expansion}). However, for the experimental detection of the random field one needs to calculate $B^z_{eff}$ beyond the perturbative regime.
We therefore calculate $B^z_{eff}$ as function of $B^x$ using an exact numerical diagonalization of a \textit{full} (18496 X 18496) two Ho ions Hamiltonian:

%%%%%%%%%%%%%%%%%%%%%%%%%%%%%%%%%%%%%%%%%%%%%%%%%%%%%%%%%%%%%%%%%%%%%%%%%%%%%
%Equation: two ions Hamiltonian
\[H_{2Ho}=H_{CF}+H_{HF}+H_{dipolar}+H_{zeeman}\]
%%%%%%%%%%%%%%%%%%%%%%%%%%%%%%%%%%%%%%%%%%%%%%%%%%%%%%%%%%%%%%%%%%%%%%%%%%%%%
with an applied transverse magnetic field along one of the crystallographic "hard" axes chosen as the $x$ axis.

\begin{figure}[h]
  \centering
  \includegraphics[width=0.5\textwidth]{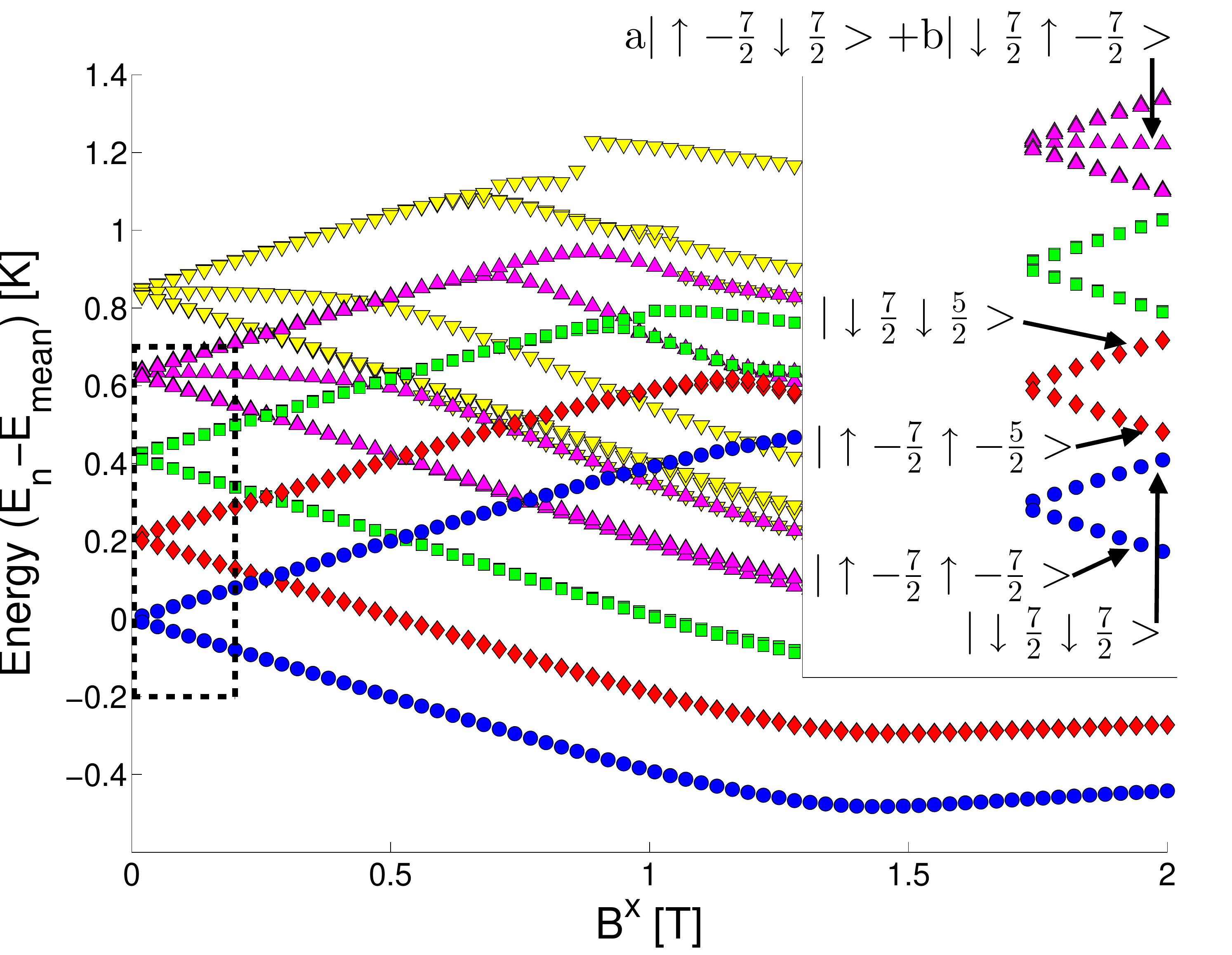}
  \caption{Some of the energy levels (lowest 32 eigenvalues) as a function of transverse field for n.n. pairs at relative positions ${\vec{r}=(\frac{1}{2} a,0,\frac{1}{4} c)}$. The inset shows a close up of the area delimited by a dashed line with some of the states annotated. The different colors and marker styles distinguish between diabatic states (see App. \ref{numerics}.) with states of similar energy at zero field sharing the same color and marker style. Zero energy is continuously calibrated to be at the mean between the two blue circles levels (which correspond to  ${|\uparrow -\frac{7}{2} \uparrow -\frac{7}{2}>}$ and ${|\downarrow \frac{7}{2} \downarrow \frac{7}{2}>}$ at zero field).}
  \label{fig:numerical_energy_levels_nn}
\end{figure}

We exploit the sparsity of the Hamiltonian matrix and use the iterative Arnoldi process \cite{Arnoldi_complexity} (see App. \ref{numerics}). The pair Hamiltonian is diagonalized and the effective longitudinal field is calculated for pairs at various nearby relative positions.
In Fig. \ref{fig:numerical_energy_levels_nn}. we plot the lowest energy levels for transverse fields between $0$ and $2$T for a n.n. pair (see also Fig. \ref{fig:LihoF4_cell}).

%%%%%%%%%%%%%%%%%%%%%%%%%%%%%%%%%%%%%%%%%%%%%%%%%%%%%%%%%%%%%%%%%%%%%%%%%%%%%
%Figure: Numerical Energy levels (n.n.)
%%%%%%%%%%%%%%%%%%%%%%%%%%%%%%%%%%%%%%%%%%%%%%%%%%%%%%%%%%%%%%%%%%%%%%%%%%%%%

The effective longitudinal field for such a pair was determined (as in eq. \ref{eq:effective_field}) by the energy difference between the ${|\uparrow -\frac{7}{2} \uparrow -\frac{7}{2}>}$ and ${|\downarrow \frac{7}{2} \downarrow \frac{7}{2}>}$ levels (blue circles in Fig. \ref{fig:numerical_energy_levels_nn}).
%%%%%%%%%%%%%%%%%%%%%%%%%%%%%%%%%%%%%%%%%%%%%%%%%%%%%%%%%%%%%%%%%%%%%%%%%%%%%
%Figure: Numerical Energy Difference
\begin{figure}[h]
  \centering
  \includegraphics[width=0.3\textwidth]{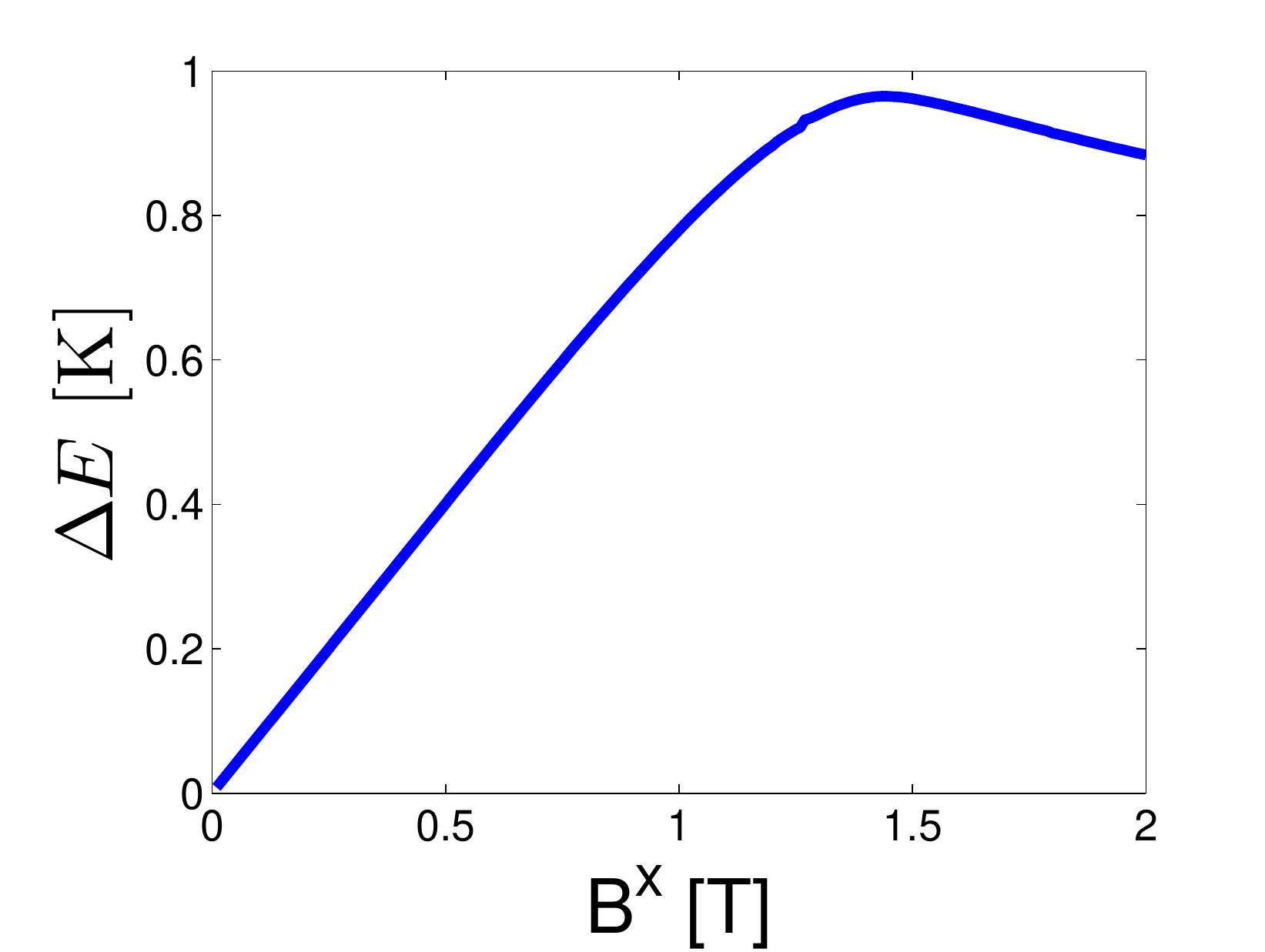}
  \caption{Energy difference between the  ${|\uparrow -\frac{7}{2} \uparrow -\frac{7}{2}>}$ and ${|\downarrow \frac{7}{2} \downarrow \frac{7}{2}>}$ states for n.n. pairs.}
  \label{fig:Energy_difference_nn}
\end{figure}
%%%%%%%%%%%%%%%%%%%%%%%%%%%%%%%%%%%%%%%%%%%%%%%%%%%%%%%%%%%%%%%%%%%%%%%%%%%%%
This energy difference, shown in Fig. \ref{fig:Energy_difference_nn},  turns out to be practically linear up to transverse field values of above $1$T, i.e. well beyond the perturbative regime. The function $\Delta E=[0.81 B^x -0.01 (B^x)^3]$K fits this curve well up to $B^x \approx 1.2$T. The linear coefficient is in excellent agreement with the perturbative expansion $\delta E_{perturbative}=0.8135 B^x$K (see App. \ref{Perturbative Expansion}). The corresponding effective field for n.n. pairs is therefore ${B^z_{eff}=0.044B^x}$.

\begin{figure}[h]
  \centering
  \includegraphics[width=0.5\textwidth]{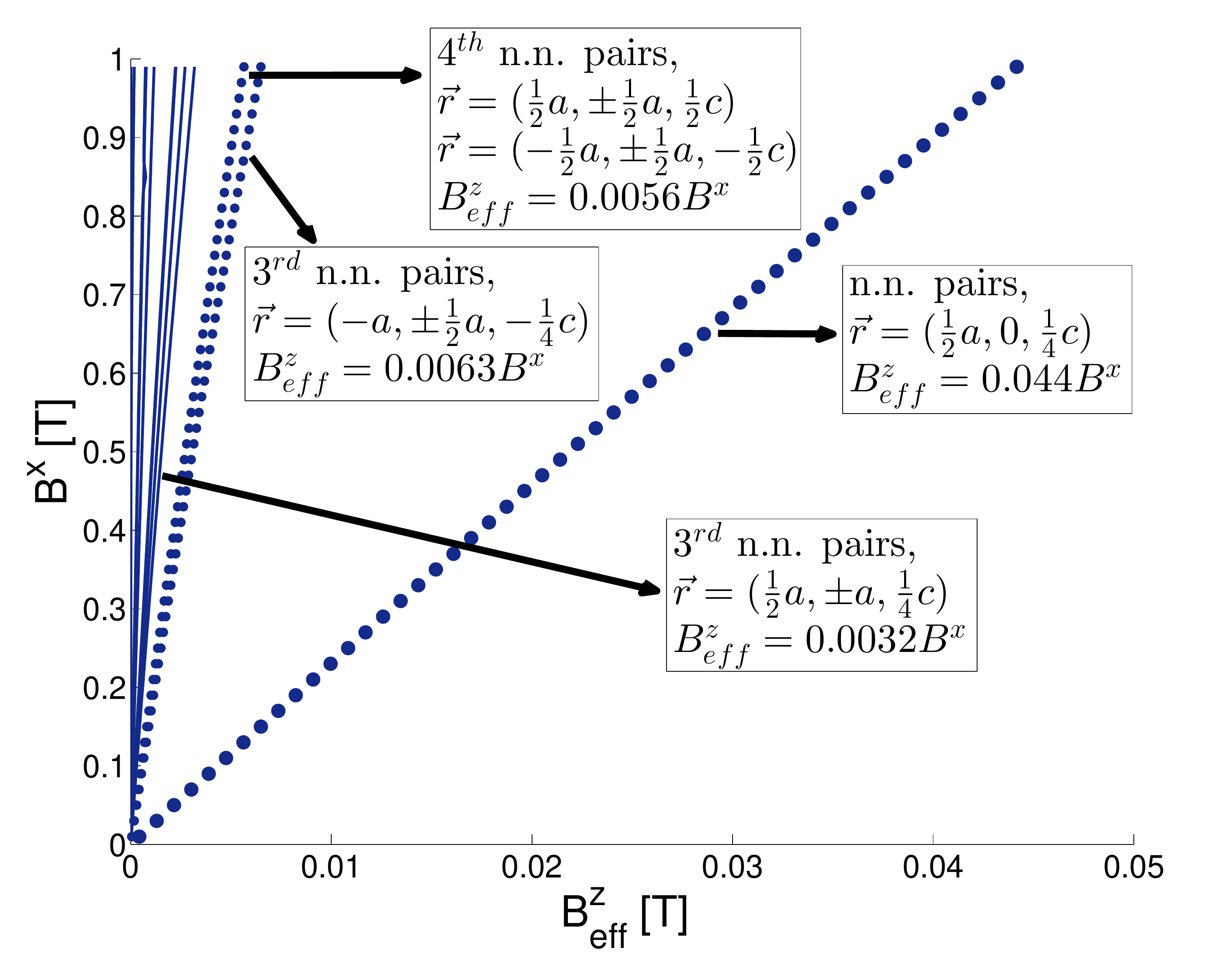}
  \caption{Effective longitudinal fields for lattice neighbors of the 8 shortest distances (The axes are switched to help resolve similar curves from one another). The ${\text{2}^{\text{nd}}}$ and ${\text{5}^{\text{th}}}$ n.n. pairs are not shown since they lie on the $XY$ plane (e.g. ${\vec{r}=(\pm a,0,0)}$) and therefore do not experience an effective field which requires finite off diagonal dipolar terms .The 4 strongest effective longitudinal fields, which we hope to measure in experiment, are accentuated. Only pairs producing positive effective fields are shown. For each pair plotted there is a conjugate pair which sees a $B^z_{eff}$ of the same absolute magnitude and opposite sign so that the negative side of the $B^z_{eff}$ axis is simply a mirror image of this figure. The relative positions of these conjugate pairs can be obtained by changing the sign of the x component in the positions shown.}
  \label{fig:Random_Fields}
\end{figure}

This linearity,  below $B^x\approx 1$T, of the energy difference is general for all pairs (both FM and AFM) and Fig. \ref{fig:Random_Fields} shows a unified plot of the effective fields for various pairs. A generalization to any other direction of the applied field could be easily obtained. At this point we wish to iterate that the effective fields in the system are determined by the random distribution of the Ho pairs and it is in this sense that the field is random.
%%%%%%%%%%%%%%%%%%%%%%%%%%%%%%%%%%%%%%%%%%%%%%%%%%%%%%%%%%%%%%%%%%%%%%%%%%%%%
%Figure: Various random fields
%%%%%%%%%%%%%%%%%%%%%%%%%%%%%%%%%%%%%%%%%%%%%%%%%%%%%%%%%%%%%%%%%%%%%%%%%%%%%

\section{Measuring the random field}
\label{Measurement}

In Sec.\ref{Random Effective Fields} we have shown that in the presence of an applied transverse magnetic field, each pair of spatially nearby spins experiences an effective longitudinal field which is specific to the relative positions of the two Ho ions. These effective fields will show as shifts in the susceptibility profile as a function of an applied longitudinal magnetic field $B^z$ and under the right protocol as distinct susceptibility peaks. We first discuss such an experimental protocol in non-equilibrium. This protocol follows the experiments of Giraud {\it et. al.}\cite{Giraud2001,Giraud2003}, only with an additional applied transverse magnetic field. We then consider similar experiments in equilibrium. We calculate explicitly the magnetization and susceptibility curves, and suggest specific parameters which are favorable for the detection of the effective longitudinal field.

\subsection{Resonances and Hysteresis of Susceptibility}
\begin{figure}[h]
  \centering
  \includegraphics[width=0.4\textwidth]{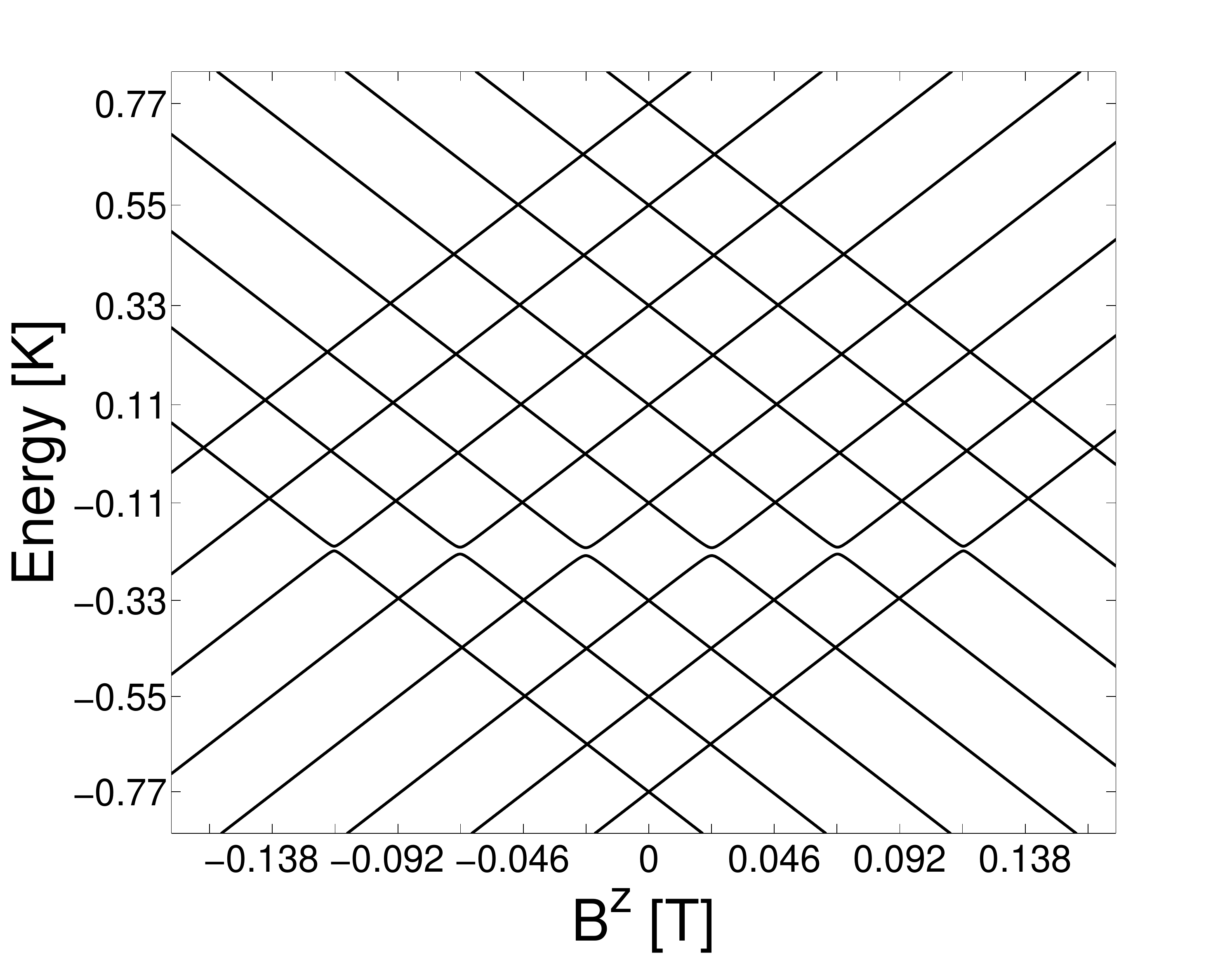}
  \caption{Energy levels of the 8 X 2 lowest electro-nuclear levels as a function of an applied longitudinal magnetic field $B^z$ (single ion picture). $B^x=0$. The level crossings are practically equally separated in the horizontal axis (${\Delta B^z=23 }$mT). This figure based on a similar plot by Giraud {\it et. al.} \cite{Giraud2001} }
\label{fig:HF_Splitting}
\end{figure}

%Giraud's experiment
In a strong dilution of the Ho ions, a first approximation would be to treat these ions singly, neglecting the inter-ionic interaction completely. In the single ion picture, an applied longitudinal field would shift the HF levels to produce the energy spectrum as in Fig. \ref{fig:HF_Splitting}.
%%%%%%%%%%%%%%%%%%%%%%%%%%%%%%%%%%%%%%%%%%%%%%%%%%%%%%%%%%%%%%%%%%%%%%%%%%%%%
%Figure: Analytical SINGLE ion Energy Levels with Longitudinal Field
%%%%%%%%%%%%%%%%%%%%%%%%%%%%%%%%%%%%%%%%%%%%%%%%%%%%%%%%%%%%%%%%%%%%%%%%%%%%%
A back and forth sweep of the longitudinal field (non-equilibrium) applied to the $\LHx$ compound at low temperatures, is expected to produce tunnelling of ions at resonant field values corresponding to the crossings in Fig. \ref{fig:HF_Splitting}. This was indeed observed in the original experiments as steps in the magnetization hysteresis curve and peaks in the corresponding susceptibility (Fig. 3 in Ref.~\onlinecite{Giraud2001}). Each of these observed resonances was labelled with an integer number $n$ corresponding to the distance from zero field in units of $\Delta B^z=23$mT.

\begin{figure}[h]
  \centering
  \includegraphics[width=0.5\textwidth]{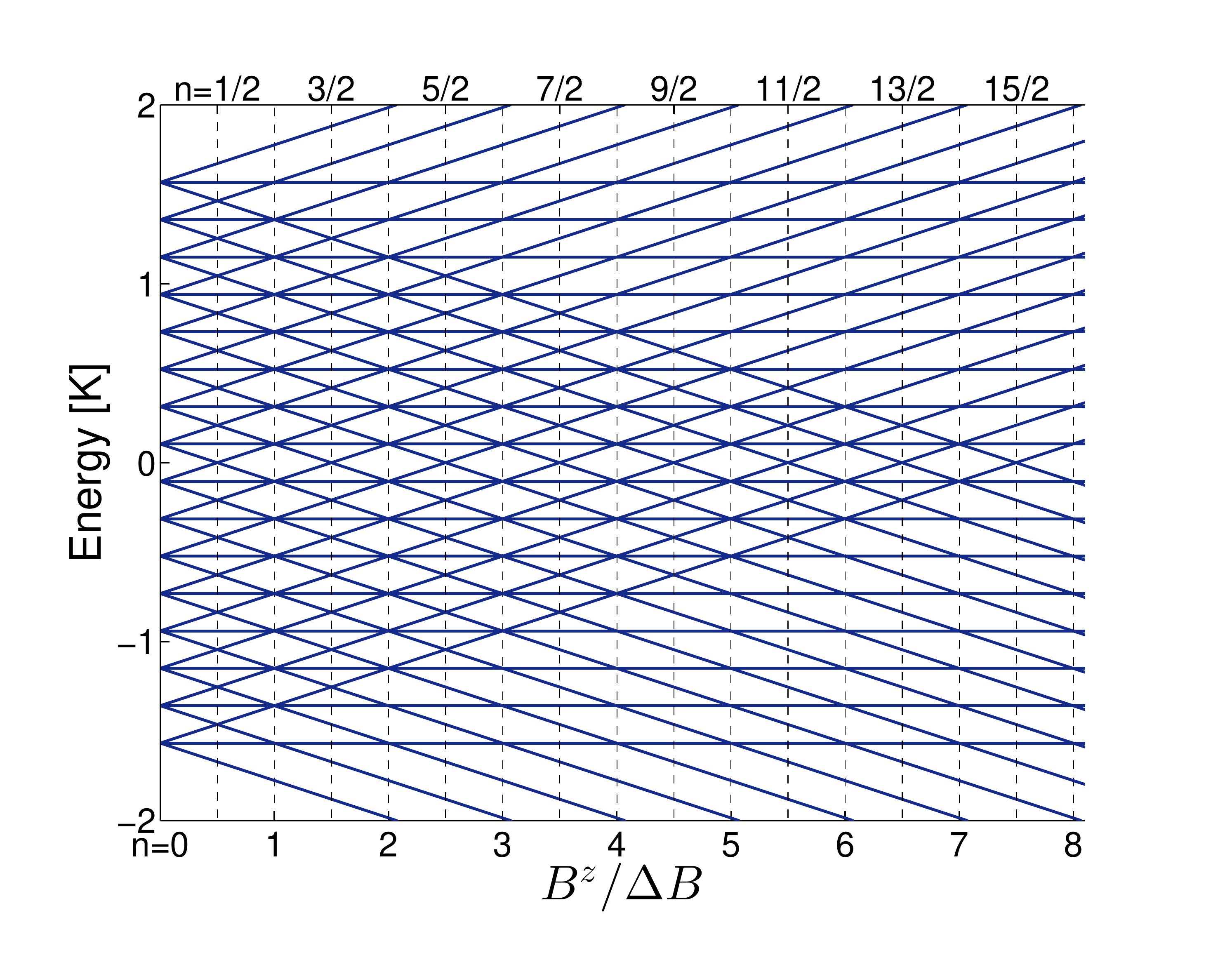}
  \caption{Energy levels in the two ion picture of the (8 X 2)$^{2}$ lowest electro-nuclear levels (i.e. all the levels corresponding to both ions being either ${|\uparrow>}$ or ${|\downarrow>}$) as a function of an applied longitudinal magnetic field. No dipolar interaction. The field is given in units of the distance between single-flip resonances $\Delta B^z=23 $mT }
  \label{fig:energy_levels_two_ion_picture_no_dipolar}
\end{figure}

%Giraud's experiment - high weep rate
However a similar experiment at a higher sweep rate (Fig. 5 in Ref. \onlinecite{Giraud2001}) showed additional tunneling resonances exactly midway between the original resonances (half integer $n$).
These were explained as the outcome of co-tunneling of two ions. The extension of Fig.\ref{fig:HF_Splitting} to a two ions picture shows these extra resonances clearly (Fig. \ref{fig:energy_levels_two_ion_picture_no_dipolar}).
In the Hilbert space of two ions, there are two Zeeman terms, so that the slope with field can be either double the slope of single ions (for two aligned spins ${|\downarrow\downarrow>}$ or ${|\uparrow\uparrow>}$) or practically zero (for anti-aligned (A-A) spins ${\alpha | \uparrow \downarrow>+ \beta | \downarrow\uparrow>}$.
%%%%%%%%%%%%%%%%%%%%%%%%%%%%%%%%%%%%%%%%%%%%%%%%%%%%%%%%%%%%%%%%%%%%%%%%%%%%%
%Figure: Analytical TWO ion Energy Levels with Longitudinal Field
%%%%%%%%%%%%%%%%%%%%%%%%%%%%%%%%%%%%%%%%%%%%%%%%%%%%%%%%%%%%%%%%%%%%%%%%%%%%%
Fig. \ref{fig:energy_levels_two_ion_picture_no_dipolar} also explains the susceptibility peaks observed by Giraud et. al. \cite{Giraud2001} at large integer ${n>7}$ through the additional crossings witnessed for the appropriate field values.

%%%%%%%%%%%%%%%%%%%%%%%%%%%%%%%%%%%%%%%%%%%%%%%%%%%%%%%%%%%%%%%%%%%%%%%%%%%%%
%proposed direct measurement
\subsection{Hysteresis under a transverse field}
\label{Hysteresis}
Repeating the non-equilibrium experiments of Giraud {\it et. al.} with the addition of a constant applied transverse field would shift the resonant field values for each ion pair by the effective fields we calculate in eq. \ref{eq:effective_field}. The contribution of each pair to the susceptibility would be shifted by a different amount, thereby generating, in principle, more peaks. For these additional peaks to be observed they have to lie well outside the intrinsic dipolar broadening of the primary peaks. Within the afore-mentioned limit of ${|B^x|<1.2T}$, and given the width of the primary resonances at $T=10$mK, only the ${\text{1}^{\text{st}}}$, ${\text{3}^{\text{rd}}}$ and ${\text{4}^{\text{th}}}$ n.n. pairs can produce adequately shifted peaks, see relative positions and specific field values in Fig. \ref{fig:Random_Fields}. Since these shifted peaks are generated only by specific pairs, their magnitude is small. Thus, their detection relies on a careful choice of parameters that places the pair peaks at field values in between unshifted primary peaks, and on following their shift as function of transverse field.

\begin{figure}[h]
	\centering
	\subfiguretopcaptrue
		\subfigure[ ${B^x=72 }$mT]{
             \includegraphics[width=0.5\textwidth]{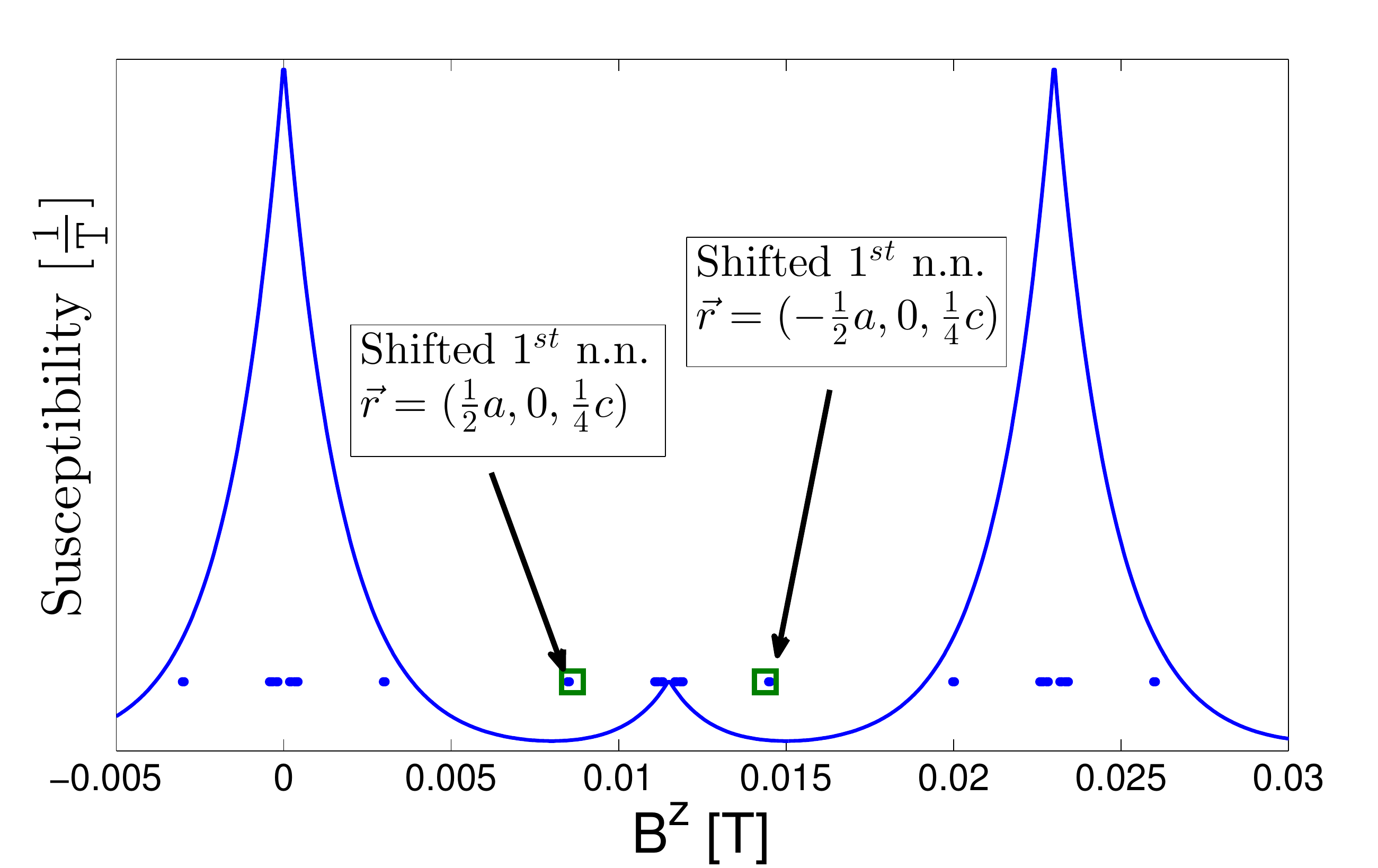}
  				\label{fig:manipulate_72mT}
  				}\\
       \subfigure[ ${B^x=464 }$mT]{
             \includegraphics[width=0.5\textwidth]{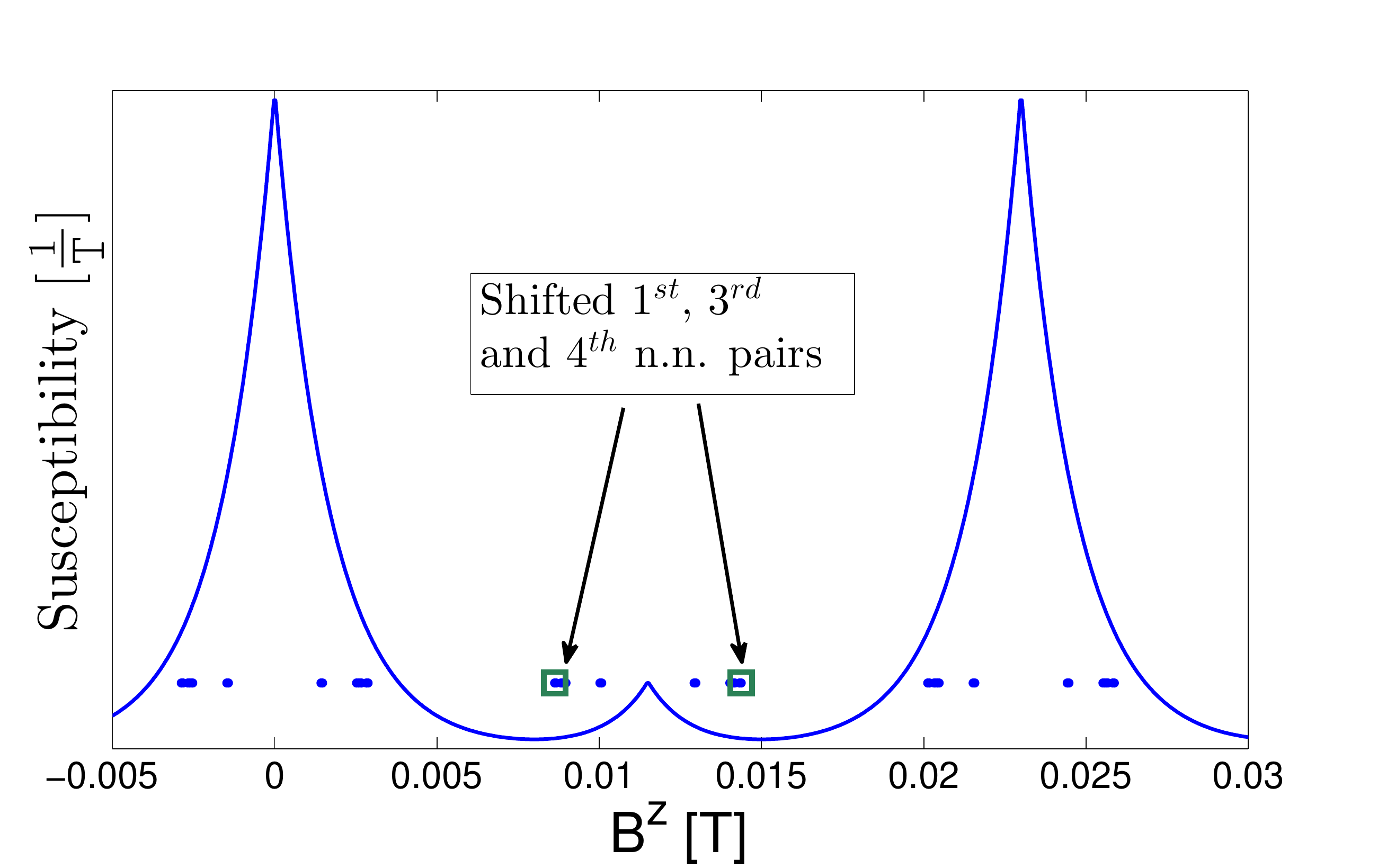}
 				 \label{fig:manipulate_464mT}
 				 }
            \caption{Predicted $B^z$ values (blue dots) of shifted susceptibility ($\chi^{zz}$) peaks for two transverse field values. a)${B^x=72 }$mT b)${B^x=464 }$mT. Blue solid lines \textit{illustrate} the unshifted primary peaks for perspective only (different functional form than experimental results but similar peak width). Green squares mark (experimentally obtained\cite{Giraud2003}) valley minima. The shifted peaks which are expected to be clearly visible in experiment are noted. All other shifted peaks (blue dots) are expected to broaden the unshifted primary peaks, without being separately visible. The field value of (b) was specifically chosen so that the contributions of ${\text{1}^{\text{st}}}$, ${\text{3}^{\text{rd}}}$ and ${\text{4}^{\text{th}}}$ n.n. pairs would coincide and combine to make a more distinct peak in such an experiment}
        \label{fig:manipulate_both_Bx}
\end{figure}

Examples of predicted positions of shifted peaks are shown in Figs. \ref{fig:manipulate_72mT} and \ref{fig:manipulate_464mT}.
%%%%%%%%%%%%%%%%%%%%%%%%%%%%%%%%%%%%%%%%%%%%%%%%%%%%%%%%%%%%%%%%%%%%%%%%%%%%%
%Figure: Dynamical sweep 72mT and 464mT
%%%%%%%%%%%%%%%%%%%%%%%%%%%%%%%%%%%%%%%%%%%%%%%%%%%%%%%%%%%%%%%%%%%%%%%%%%%%%%%%
The shifted peaks belonging to the ${\text{3}^{\text{rd}}}$ and ${\text{4}^{\text{th}}}$ n.n. pairs (see Fig. \ref{fig:manipulate_464mT}) can only become discernible for ${B^x \geq 0.4}$T.  The ability to precisely predict the position of these peaks is a direct consequence of our exact calculation of the effective longitudinal field beyond perturbation theory regime.
\par
This experimental protocol could prove the validity of the ansatz employed by Giraud at al. \cite{Giraud2001} which explains the small susceptibility peaks at half integer $n$ as co-tunnelling peaks. Furthermore the size of the shifted peaks should indicate the relative contribution of each pair to the co-tunnelling peaks in the original experiment. Another advantage of this protocol is the ability to study the combined effects of interaction, quantum fluctuations induced by the transverse field, and the effective longitudinal field on the dynamics of the system.
However the above method has the disadvantage of producing results for which the distinctness of the pair resonances against the 'background' of the unshifted resonances is highly sensitive to experimental parameters (such as transverse field, sweep rate, and measurement resolution).
We now discuss a variation of this experiment where the applied field is swept adiabatically so that the system stays at instantaneous equilibrium throughout the sweep.

%%%%%%%%%%%%%%%%%%%%%%%%%%%%%%%%%%%%%%%%%%%%%%%%%%%%%%%%%%%%%%%%%%%%%%%%%%%%%
\subsection{Adiabatic field sweep}
An adiabatic sweep leads to a Boltzmann population of states where for temperatures much lower than the $215$mK HF splitting, only the instantaneous ground state is significantly populated.
%%%%%%%%%%%%%%%%%%%%%%%%%%%%%%%%%%%%%%%%%%%%%%%%%%%%%%%%%%%%%%%%%%%%%%%%%%%%%
%Figure: energy levels- ground crossings
\begin{figure}[h]
	\centering
		\subfiguretopcaptrue
		\subfigure[$1^{st}$ n.n. pairs (FM)]{
             \includegraphics[width=0.4\textwidth]{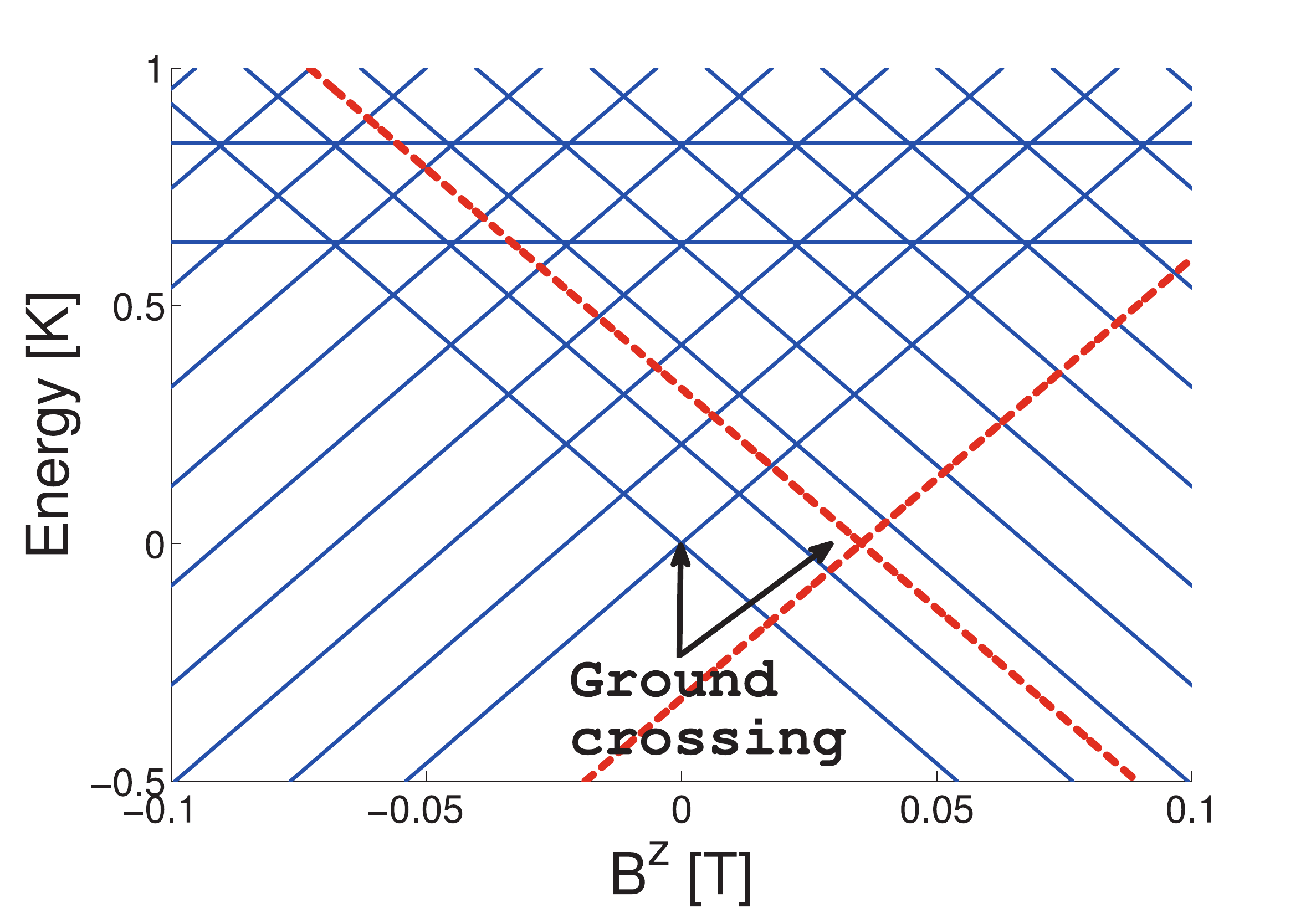}
  				\label{fig:illustrated_energy_nn}
  				}
       \subfigure[$3^{rd}$ n.n. pairs (AFM)]{
             \includegraphics[width=0.4\textwidth]{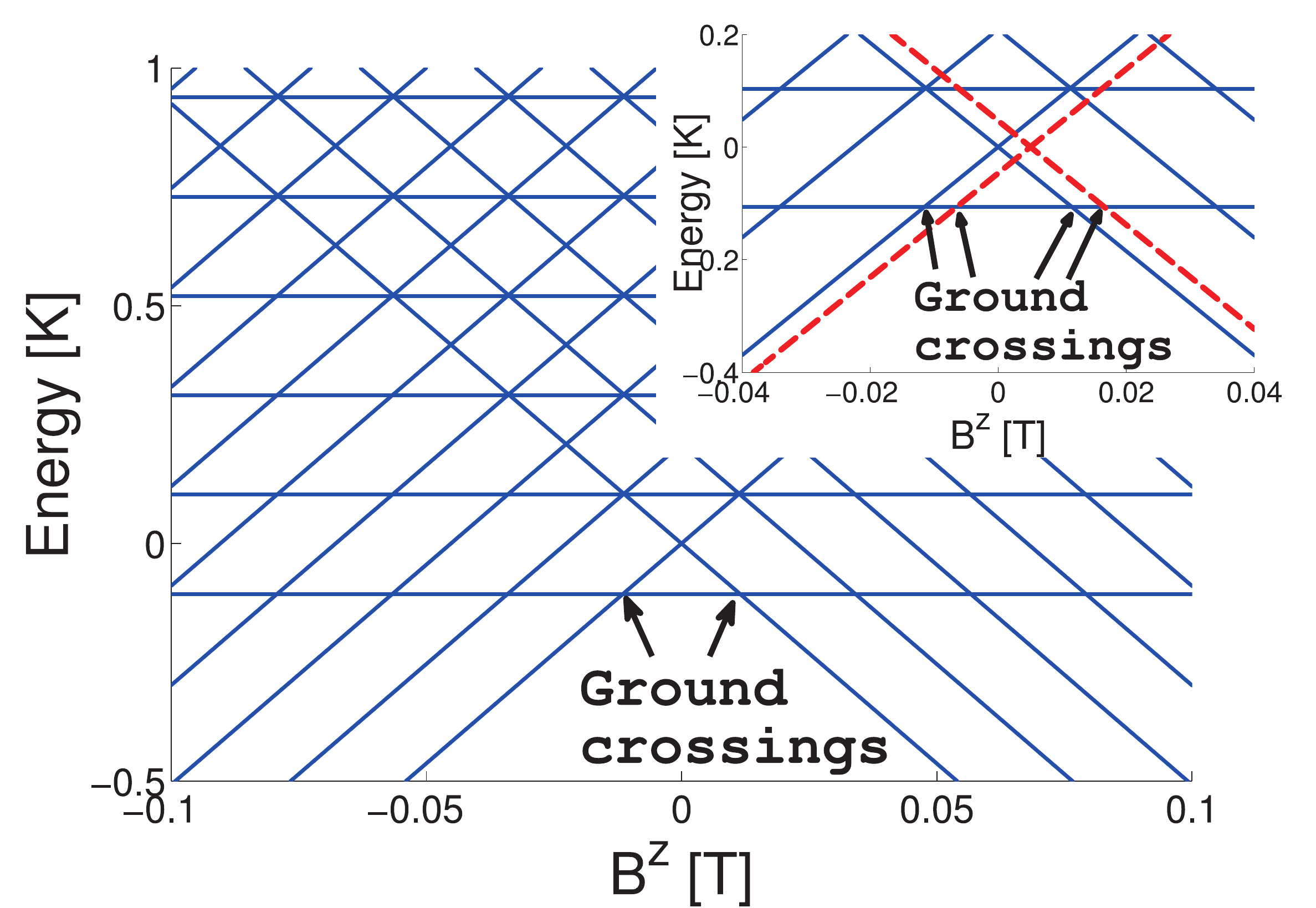}
 				 \label{fig:illustrated_energy_2nd_nn}
 				 }
        \caption{Energy levels as a function of $B^z$ in zero transverse field (solid blue) for FM (${\text{1}^{\text{st}}}$ n.n.) and AFM (${\text{3}^{\text{rd}}}$ n.n.) pairs. The resonances involving ground levels are the only ones that will show in an adiabatic sweep at a very low temperature. The dashed red lines show the shift in ground states energies due to an effective field (${B^z_{eff}=35}$mT for $1^{st}$ n.n. and ${B^z_{eff}=5}$mT for $3^{rd}$ n.n.) generated when a transverse field ${B^x=0.8T}$ is applied. The relevant resonances in (b) are magnified in the inset for a better view of the energy shift due to the weak effective field.}
        \label{fig:energy_levels_two_ion_picture_FM_and_AFM}
\end{figure}
%%%%%%%%%%%%%%%%%%%%%%%%%%%%%%%%%%%%%%%%%%%%%%%%%%%%%%%%%%%%%%%%%%%%%%%%%%%%%
This is in contrast to the situation in Sec. \ref{Hysteresis}, where excited states could be significantly populated due to the finite sweep rate. For pairs of spins, the instantaneous ground states depend on the intra-pair dipolar interaction, see representative examples in Fig. \ref{fig:energy_levels_two_ion_picture_FM_and_AFM}.

Let us consider first the case of $B^x=0$. For FM pairs the {A-A spins state is located at a higher energy, and the ground states change directly from the state ${| \uparrow -\frac{7}{2} \uparrow -\frac{7}{2}>}$ to the state ${| \downarrow \frac{7}{2} \downarrow \frac{7}{2}>}$ at $B^z=0$. For AFM pairs the {A-A spins state is at a lower energy. Sweeping the longitudinal field, e.g. from positive to negative, these pairs start at the ${| \uparrow -\frac{7}{2} \uparrow -\frac{7}{2}>}$ state, which first changes to the A-A spins state at $B^z>0$ and to the ${| \downarrow \frac{7}{2} \downarrow \frac{7}{2}>}$ state at $B^z<0$. If we now introduce a finite $B^x$, the energy of the ${| \uparrow -\frac{7}{2} \uparrow -\frac{7}{2}>}$ state increases while that of ${| \downarrow \frac{7}{2} \downarrow \frac{7}{2}>}$ decreases. The energy of the A-A spins state stays fixed. For the FM pairs this results in a linear dependence in $B^x$ of the field $B^z$ where the pair flips (the intersection of these levels moves to more positive or more negative $B^z$, depending on the intrapair orientation and the resulting sign of the effective longitudinal field). Similarly, for the AFM pairs it results in a linear shift of the values of the single spin flips.
\par
Experimentally, the above shifts in energy crossings will present themselves in the magnetization curves once the longitudinal field is swept (adiabatically).
We therefore calculate the magnetization and susceptibility as a function of $B^x$, $B^z$, temperature, and Ho concentration. As shown in the example of Fig. \ref{fig:illustrated_energy_2nd_nn}, the AFM pairs should, unrelated to the transverse or effective fields, produce small peaks adjacent to the zero $B^z$ primary peak. To make sure these resonances do not obscure the shifted peaks of the effective fields, we have explicitly calculated the susceptibility due to nearby AFM pairs (up to ${\text{5}^{\text{th}}}$ n.n.) with farther pairs included implicitly in the broadening of the central peak. The effective fields taken into account in the calculation of the susceptibility include only nearby pairs (up to ${\text{4}^{\text{th}}}$ n.n.), as farther pairs produce only negligible fields.

%%%%%%%%%%%%%%%%%%%%%%%%%%%%%%%%%%%%%%%%%%%%%%%%%%%%%%%%%%%%%%%%%%%%%%%%%%%%%
%Figure :Predicted "3D" Susceptibility
\begin{figure}[h]
\centering
 \includegraphics[width=0.5\textwidth]{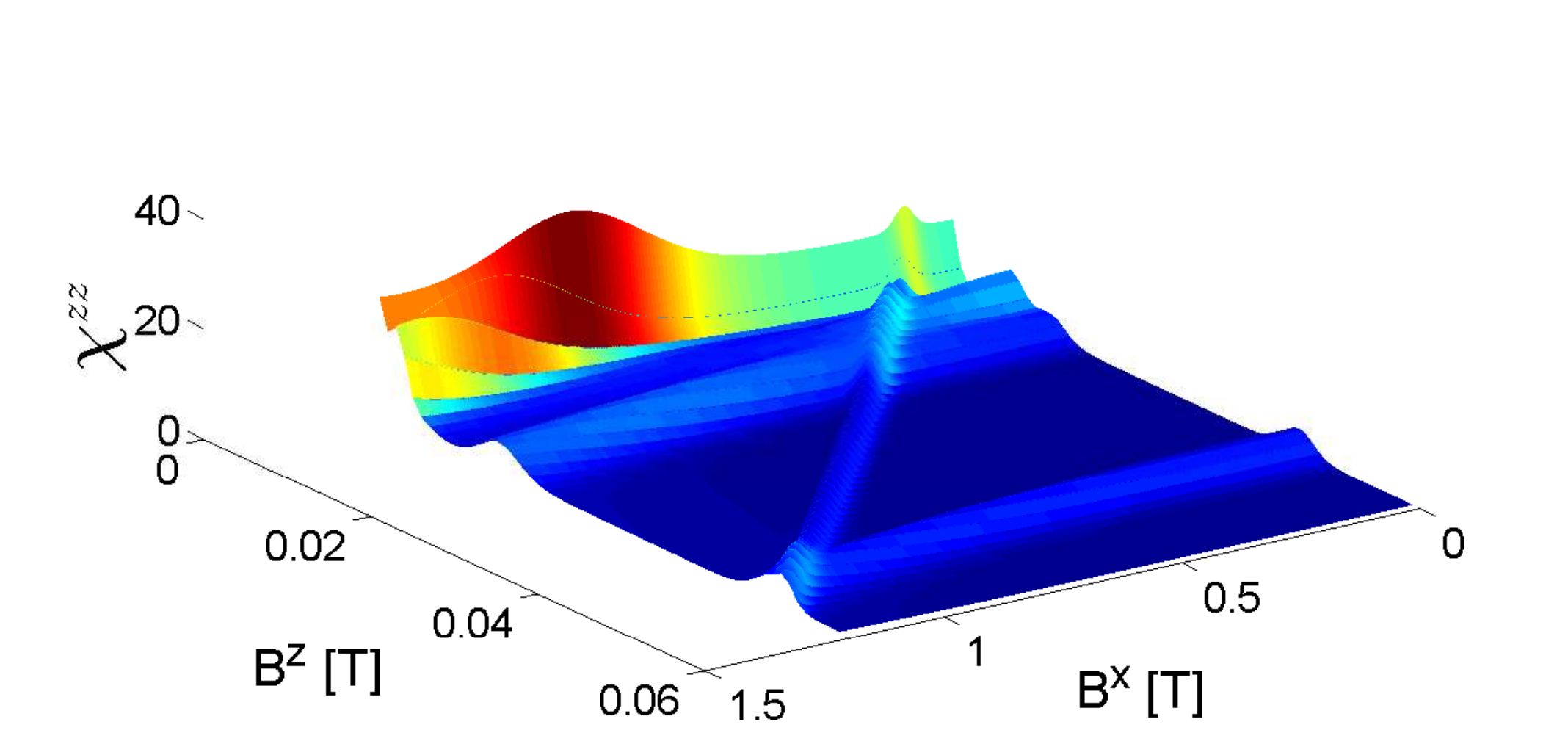}
  \caption{Predicted Susceptibility to $B^z$ (${\chi^{zz}=\frac{\partial M^z(B^x,B^z)}{\partial B^z}}$).
The large peak (at $B^z<0.004$T) is truncated for better contrast away from it.}
 \label{fig:sus_equi_3D_Jz_varies}
\end{figure}
%%%%%%%%%%%%%%%%%%%%%%%%%%%%%%%%%%%%%%%%%%%%%%%%%%%%%%%%%%%%%%%%%%%%%%%%%%%%%
%%%%%%%%%%%%%%%%%%%%%%%%%%%%%%%%%%%%%%%%%%%%%%%%%%%%%%%%%%%%%%%%%%%%%%%%%%%%%
\begin{figure}[h]
%Figure: Predicted "2D" projection Susceptibility
  \centering
  \includegraphics[width=0.5\textwidth]{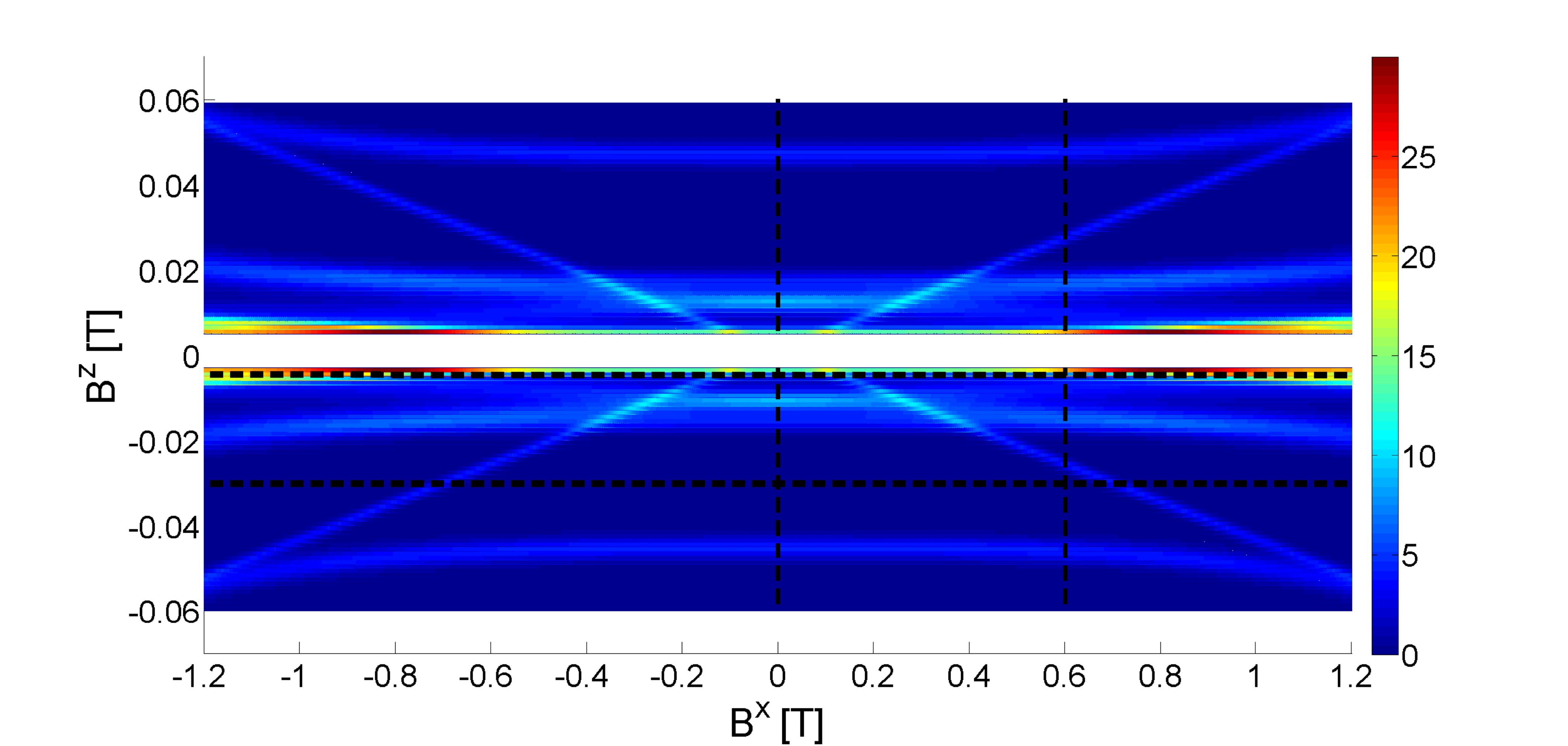}
          \caption{Projection of the predicted $\chi^{zz}$ susceptibility (color scale) on the entire ${B^xB^z}$ plane. The area of the big peak at zero longitudinal field ($-0.004<B^z<0.004$T) is omitted for better contrast away from it. The dashed lines indicate the paths used to produce Figs. \ref{fig:sus_equi_trans_field_600mT_zoom} through \ref{fig:sus_zx_equi_longi_field_4mT_zoom}.}
  \label{fig:sus_equi_3D_Jz_varies-projection}
\end{figure}
%%%%%%%%%%%%%%%%%%%%%%%%%%%%%%%%%%%%%%%%%%%%%%%%%%%%%%%%%%%%%%%%%%%%%%%%%%%%%

In Figs. \ref{fig:sus_equi_3D_Jz_varies} and \ref{fig:sus_equi_3D_Jz_varies-projection} we plot the susceptibility as a function of $B^x$ and $B^z$ for dilution ${x=0.005}$ and temperature $T=10$mK. The central peak at $B^z < 0.004$T is omitted, for a better visualization of the shifted pair peaks.
The susceptibility peaks running diagonal in $B^x$ and $B^z$ correspond to the shifted resonances of ${\text{1}^{\text{st}}}$ n.n. pairs. The shifted peaks associated with the ${\text{3}^{\text{rd}}}$ and ${\text{4}^{\text{th}}}$ n.n. pairs are harder to discern at this resolution. Changing the concentration $x$ or temperature $T$ will not affect the peaks' positions. Both the height and width of the shifted peaks increase with $x$ while a decrease in temperature simply narrows the peaks.

Finding the effective fields therefore amounts to finding the positions of the shifted peaks in the $B^x$ $B^z$ plane. The advantage of the adiabatic protocol is that by choosing the right field sweep path in the $B^x$ $B^z$ plan, we can isolate the contribution of a shifted peak from any other susceptibility features. We demonstrate three such paths that should show distinct shifted peaks (see the dashed lines in Fig. \ref{fig:sus_equi_3D_Jz_varies-projection}).

The first path is a sweep of the longitudinal field with a constant \textit{transverse} field $B^x=0.6$T.
%%%%%%%%%%%%%%%%%%%%%%%%%%%%%%%%%%%%%%%%%%%%%%%%%%%%%%%%%%%%%%%%%%%%%%%%%%%%%
%Figure: Predicted Susceptibility Bx=600mT
\begin{figure}[h]
  \centering
  \includegraphics[width=0.4\textwidth]{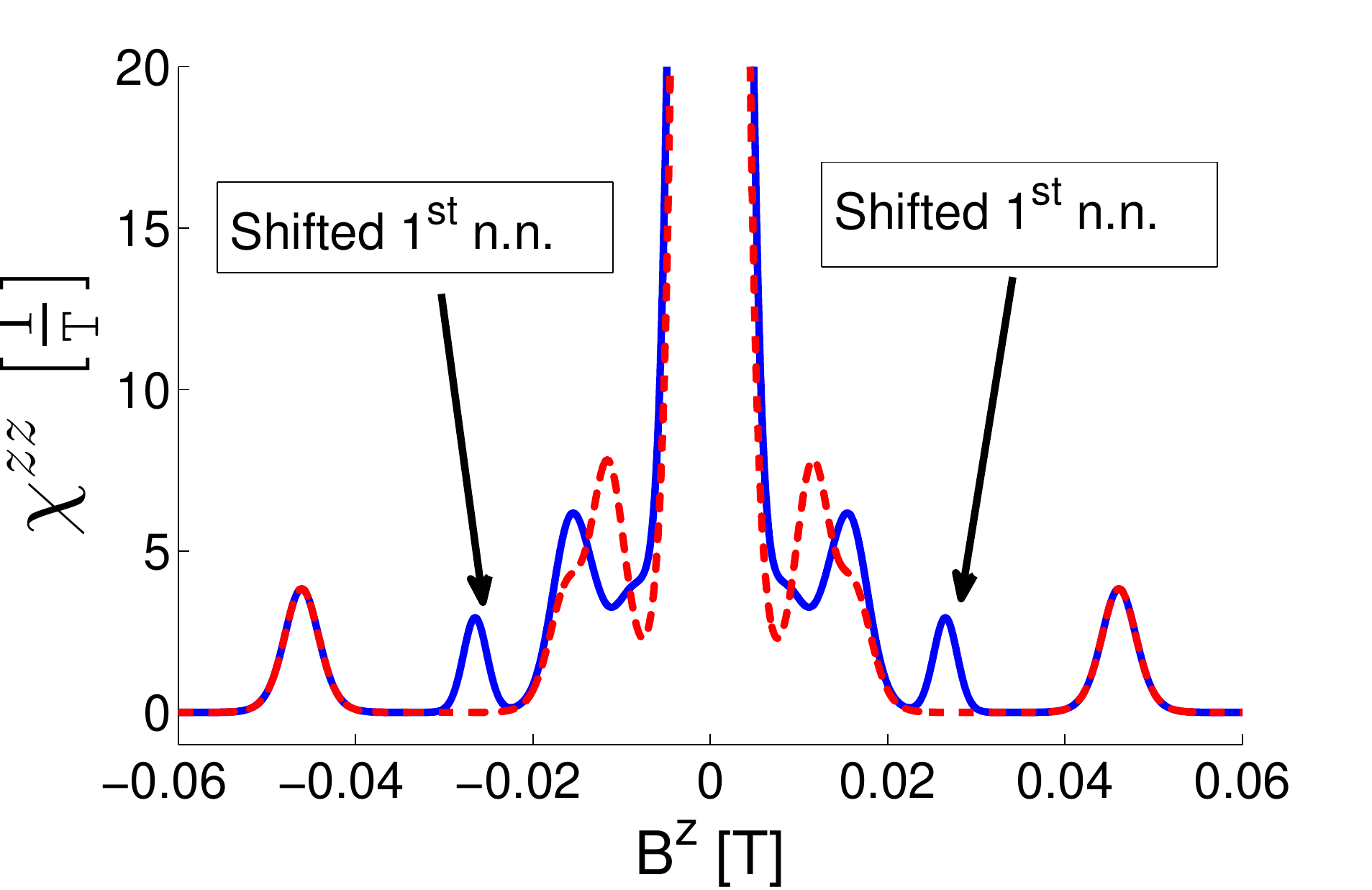}
  \caption{Predicted susceptibility for an adiabatic sweep of ${B^z}$ with constant $B^x=0.6$T (solid blue). The same is depicted for ${B^x=0}$ (dashed red) for comparison. Besides the shifted peaks clearly noted in the figure, and the big zero field peak in the center, some unshifted peaks of AFM pairs are also visible corresponding (from outside in) to the ${\text{2}^{\text{nd}}}$, ${\text{3}^{\text{rd}}}$ and ${\text{5}^{\text{th}}}$ n.n. pairs. }
  \label{fig:sus_equi_trans_field_600mT_zoom}
\end{figure}
%%%%%%%%%%%%%%%%%%%%%%%%%%%%%%%%%%%%%%%%%%%%%%%%%%%%%%%%%%%%%%%%%%%%%%%%%%%%%
Such a path shows the shifted peaks for n.n. pairs along with other unshifted peaks (Fig. \ref{fig:sus_equi_trans_field_600mT_zoom}). The latter are due to pairs experiencing much weaker or zero $B^z_{eff}$ and help to put the former in perspective. A small change in $B^x$ would change the positions of the shifted peaks linearly.

The second path is a sweep of the transverse field with a constant \textit{longitudinal} field $B^z=-30$mT. At ${B^x=0}$ all of the spins are down (except some ${\text{2}^{\text{nd}}}$ n.n. AFM pairs, and when the} transverse field is changed, only the ${\text{1}^{\text{st}}}$ n.n. pairs see a significant effective field, so that only (some of) them cross a resonance and flip up showing a change in magnetization and therefore significant susceptibility (Fig. \ref{fig:sus_equi_longi_field_30mT_zoom}). This path was selected because it shows the shifted peaks for n.n. pairs without any unshifted peaks to obscure them.

%%%%%%%%%%%%%%%%%%%%%%%%%%%%%%%%%%%%%%%%%%%%%%%%%%%%%%%%%%%%%%%%%%%%%%%%%%%%%
\begin{figure}[h]
%Figure: Predicted Susceptibility Bz=-30mT
  \centering
  \includegraphics[width=0.4\textwidth]{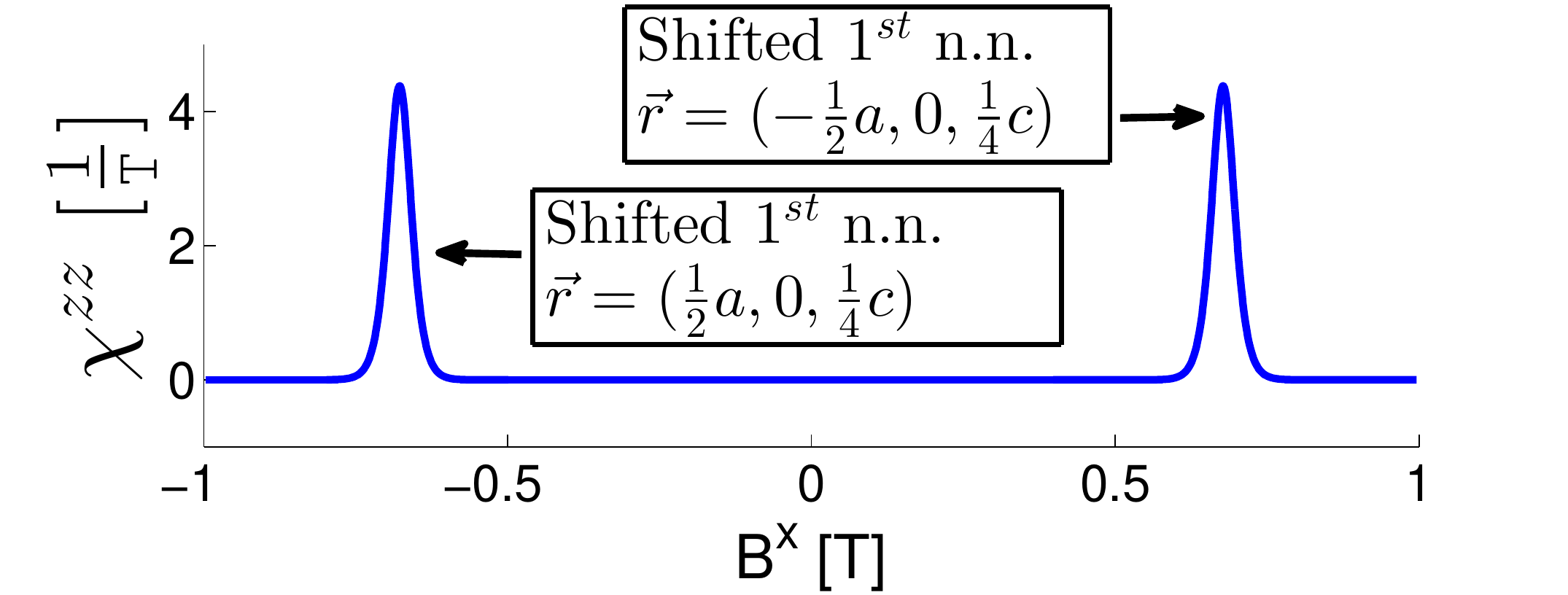}
  \caption{Predicted susceptibility (${\chi^{zz}=\frac{\partial M^z(B^x,B^z)}{\partial B^z}}$) for an adiabatic sweep of ${B^x}$ with constant $B^z=-30$mT. Each peak corresponds to one out of two different ${\text{1}^{\text{st}}}$ n.n. pairs}
  \label{fig:sus_equi_longi_field_30mT_zoom}
\end{figure}
%%%%%%%%%%%%%%%%%%%%%%%%%%%%%%%%%%%%%%%%%%%%%%%%%%%%%%%%%%%%%%%%%%%%%%%%%%%%%

The third path is again a sweep of the transverse field but with a different constant \textit{longitudinal} field $B^z=-4$mT. This path was selected since it shows the shifted peaks for both ${\text{1}^{\text{st}}}$ n.n. and ${\text{4}^{\text{th}}}$ n.n. pairs (Figs. \ref{fig:sus_zz_equi_longi_field_4mT_zoom} and \ref{fig:sus_zx_equi_longi_field_4mT_zoom}).
%%%%%%%%%%%%%%%%%%%%%%%%%%%%%%%%%%%%%%%%%%%%%%%%%%%%%%%%%%%%%%%%%%%%%%%%%%%%%
\begin{figure}[h]
%Figure: Predicted Susceptibility_zz Bz=-4mT
  \centering
  \includegraphics[width=0.4\textwidth]{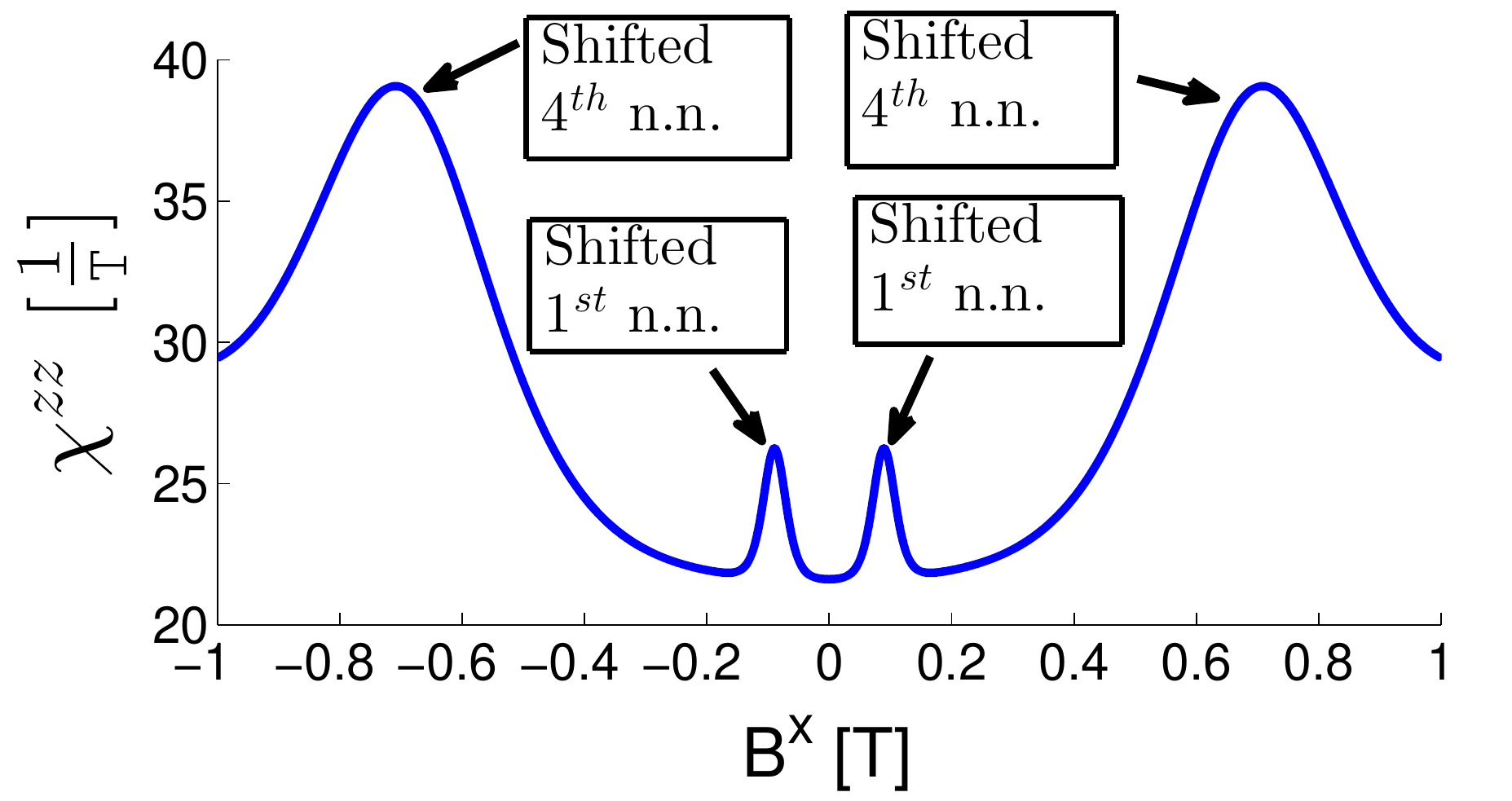}
  \caption{Predicted susceptibility (${\chi^{zz}=\frac{\partial M^z(B^x,B^z)}{\partial B^z}}$) for an adiabatic sweep of ${B^x}$ with constant $B^z=-4$mT. The peaks for the ${\text{4}^{\text{th}}}$ n.n. pairs are broader than those for the ${\text{1}^{\text{st}}}$ n.n. pairs precisely because of the lower effective field of the former. The width of the shifted peaks is in fact another measure for the effective field.}
  \label{fig:sus_zz_equi_longi_field_4mT_zoom}
\end{figure}
%%%%%%%%%%%%%%%%%%%%%%%%%%%%%%%%%%%%%%%%%%%%%%%%%%%%%%%%%%%%%%%%%%%%%%%%%%%%%
%%%%%%%%%%%%%%%%%%%%%%%%%%%%%%%%%%%%%%%%%%%%%%%%%%%%%%%%%%%%%%%%%%%%%%%%%%%%%
\begin{figure}[h]
%Figure: Predicted Susceptibility_zx Bz=-4mT
  \centering
  \includegraphics[width=0.4\textwidth]{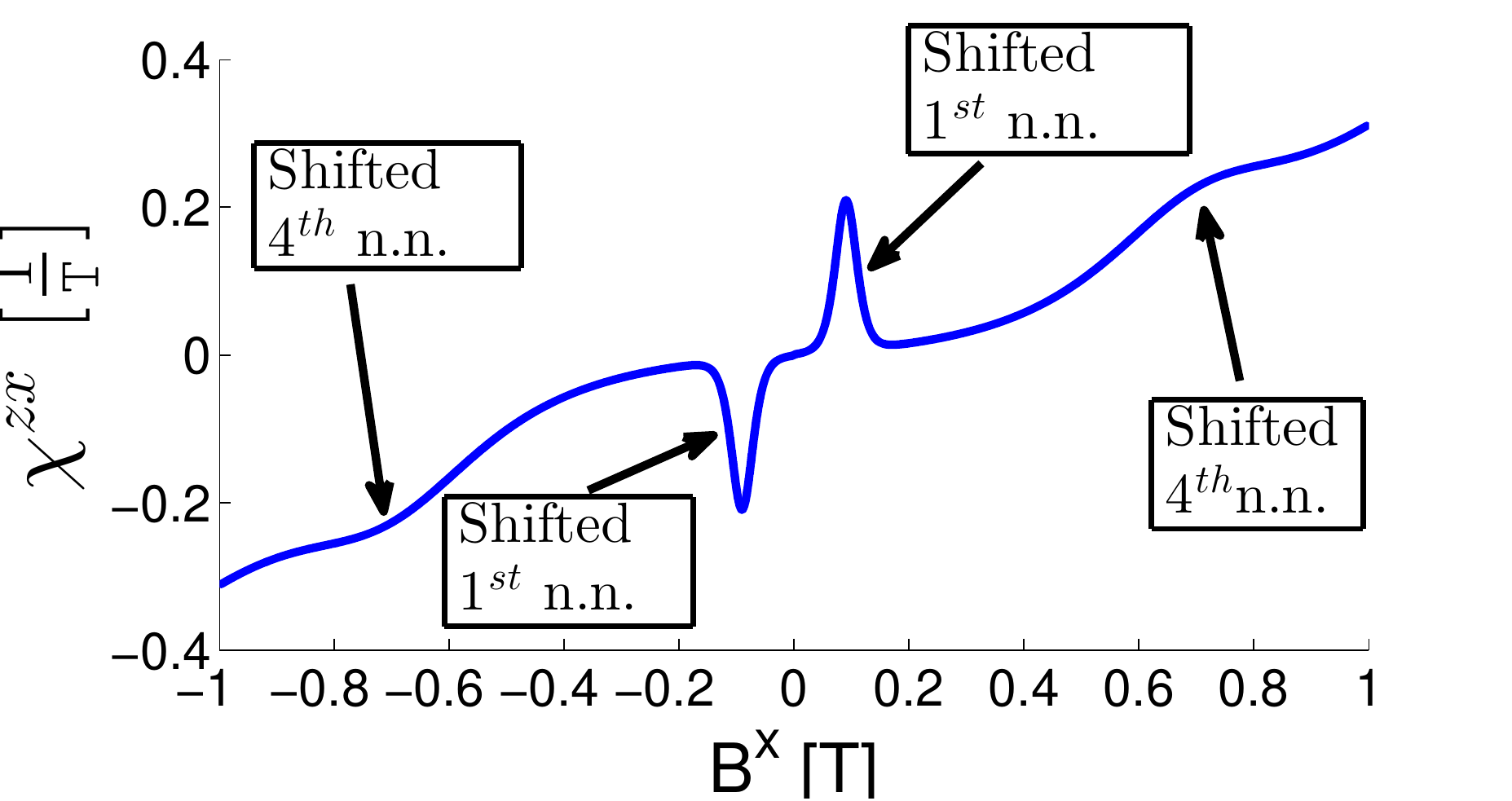}
  \caption{Predicted susceptibility (${\chi^{zx}=\frac{\partial M^z(B^x,B^z)}{\partial B^x}}$) for an adiabatic sweep of ${B^x}$ with constant $B^z=-4$mT. The reason for the inversion of the susceptibility at the left half of the figure is that for some pairs, sweeping $B^x$ positively sweeps the effective longitudinal field negatively resulting in spins flipping down instead of up.}
  \label{fig:sus_zx_equi_longi_field_4mT_zoom}
\end{figure}
%%%%%%%%%%%%%%%%%%%%%%%%%%%%%%%%%%%%%%%%%%%%%%%%%%%%%%%%%%%%%%%%%%%%%%%%%%%%%
For this path we present both the susceptibility to change in $B^z$ (${\chi^{zz}=\frac{\partial M^z(B^x,B^z)}{\partial B^z}}$) and the susceptibility to change in $B^x$ (${\chi^{zx}}$). In the latter, the different signs of the effective field realized in pairs with different intra-pair orientations are manifested not only in the positions of the susceptibility features, but also in their nature, i.e. dips vs. peaks. Note also that since
%The reasoning behind this is that
this path runs along the big (unshifted) ${\chi^{zz}}$ susceptibility peak centered around zero longitudinal field, we expect the measurement of ${\chi^{zx}}$ to be less noisy than that of ${\chi^{zz}}$. The predictions of Figs. \ref{fig:sus_equi_trans_field_600mT_zoom} through \ref{fig:sus_zx_equi_longi_field_4mT_zoom} take into account a broadening of $1mT$ due to both spin-spin interactions with farther Ho ions and HF interactions with Fluorine ions.

\section{Conclusions}
\label{conclusions}

$\LHx$ is the archetypal anisotropic dipolar magnet which constitutes a realization of the RFIM upon any finite dilution. The consequences of the random effective longitudinal field can be observed macroscopically for various Ho concentrations. In this paper we show that by considering rare pairs in very dilute $\LHx$ systems the effective longitudinal field can be readily measured. By diagonalizing exactly the full two-Ho Hamiltonian in the $\LHx$ lattice  in the presence of an applied transverse field, we calculate the effective longitudinal field experienced by various nearby Ho pairs beyond perturbation theory. We show that these effective longitudinal fields result in shifted susceptibility peaks in nonequilibrium hysteresis experiments. These shifts are nearly linear in the applied transverse field, up to a rather high field of ${B^x\approx 1.2}$T. The slope of the effective field vs. the applied transverse field depends, in both sign and magnitude, on the intra-pair distance and relative orientation. We then calculate the magnetization in equilibrium for all values of applied transverse and longitudinal magnetic fields, and deduce susceptibility curves for various paths in the $B_x - B_z$ plain. Both the sign and the magnitude of the effective longitudinal fields have clear signatures in these paths. Thus, following the protocols suggested in this paper the effective longitudinal field can be directly measured, inferring directly on the magnitude of the random field in the  $\LHx$ system at any concentration. This would provide a first direct measurement of the random field in a FM Ising-like system. In addition, such a measurement will give strong support to the conjecture that even in an extreme dilution both single spin tunneling and pair co-tunneling exist.

\begin{acknowledgments}
We would like to thank Bernard Barbara, Romain Giraud, and Nicolas Laflorencie for useful discussions. This work was supported by the Marie Curie Grant PIRG-GA-2009-256313.
\end{acknowledgments}

\appendix
\section{Perturbative Expansion}
\label{Perturbative Expansion}
The original analysis by Schechter and Laflorencie \cite{SchechterLaflorencie} employed second order perturbation theory to derive the splitting with transverse field between the degenerate Ising states of a general anisotropic dipolar magnet model Hamiltonian. Here we give the results of a similar though more detailed expansion using the specifics of the $\LHx$ compound as an indication to the validity of our numerical results (a more detailed derivation can be found in Ref. \onlinecite{YoavThesis}). Note that this perturbative analysis can include all the ions in the compound as opposed to just a pair of ions in the numerical analysis.
\par
The unperturbed Hamiltonian is composed of the CF term (Eq. \ref{eq:CF}) and the longitudinal components of the HF and dipolar interactions (Eq. \ref{eq:HF} and \ref{eq:dipolar}):
%%%%%%%%%%%%%%%%%%%%%%%%%%%%%%%%%%%%%%%%%%%%%%%%%%%%%%%%%%%%%%%%%%%%%%%%%%%%%
%Equation: H0 Unperturbed Hamiltonain
\begin{equation}
 H_0=H_{CF}+A_J\sum_i I_i^z \cdot J_i^z +\mathop{ \sum_{ i \neq j}} V_{ij}^{zz} J_i^{z} J_j^{z}
 \end{equation}
%%%%%%%%%%%%%%%%%%%%%%%%%%%%%%%%%%%%%%%%%%%%%%%%%%%%%%%%%%%%%%%%%%%%%%%%%%%%%
We are interested in the energy splitting between '\textit{${\text{Global-Ising}}$ states}' for which all the ions are in one of the (single ion) electro-nuclear Ising states (e.g. ${|\uparrow  -\frac{5}{2}>}$). Specifically the calculation is carried out for any two such states, which are degenerate and related by ${J_k^z \rightarrow -J_k^z}$ and ${I_k^z \rightarrow -{I_k^z}}$ (where $k$ is an ion index) symmetry for all ions. The ground states, which according to the scaling ("droplet") picture \cite{FisherHuse1,FisherHuse2} are only twofold degenerate, can be taken as a representative example.
\par
The perturbation $H'$ is composed of the remaining HF and dipolar components and also an applied transverse field (chosen to point along the $x$ axis):
%%%%%%%%%%%%%%%%%%%%%%%%%%%%%%%%%%%%%%%%%%%%%%%%%%%%%%%%%%%%%%%%%%%%%%%%%%%%%
%Equation: H' Perturbation
\begin{equation}
\begin{aligned}
H'= \sum_{ \alpha, \beta \neq zz} \sum_{ j \neq i} \sum_i  V_{ij}^{\alpha \beta}& J_i^{\alpha} J_j^{\beta}\\
-g_L\mu_B B^x \sum_i J_i^x +&\frac{A_J}{2} \sum_i (I_i^ +\cdot J_i^- +I_i^- \cdot J_i^+)
\end{aligned}
\end{equation}
%%%%%%%%%%%%%%%%%%%%%%%%%%%%%%%%%%%%%%%%%%%%%%%%%%%%%%%%%%%%%%%%%%%%%%%%%%%%%
We are interested here in the breaking of the degeneracy of the time reversal levels by the applied transverse field. From symmetry, only odd terms in $B^x$ appear in the perturbative expansion. In order to compare with our numerical results we therefore calculate the coefficient of the linear term. The dominant contribution comes from fluctuations to the first excited electronic state $\Gamma_2$. This contribution is given by\cite{SchechterLaflorencie}
%%%%%%%%%%%%%%%%%%%%%%%%%%%%%%%%%%%%%%%%%%%%%%%%%%%%%%%%%%%%%%%%%%%%%%%%%%%%%
%Euqation: Perturbation DeltaE leading term
\begin{equation}
\delta E=\sum_k 2 g_L\mu_B  \left[ \frac{4 \eta} {\Omega_0} \mathop{ \sum_{ i \neq k}} V_{ki}^{zx}  B^x \right] <J_k^{z}>
\label{eq:deltaE_leading_term}
\end{equation}
%%%%%%%%%%%%%%%%%%%%%%%%%%%%%%%%%%%%%%%%%%%%%%%%%%%%%%%%%%%%%%%%%%%%%%%%%%%%%
where ${\Omega_0=10.8K}$ is the energy separation between the (single ion) ground levels and the $\Gamma_2$ level. The square of the coupling between $\Gamma_2$ and the ground states ${\eta= {|<\Gamma_2 | J^x  |\uparrow>|}^2=5.62}$ is found numerically.\\
This energy splitting has the form of a sum of (single ion) longitudinal Zeeman splittings (see eq. \ref{eq:longi_deltaE})
and dictates the effective longitudinal fields
%%%%%%%%%%%%%%%%%%%%%%%%%%%%%%%%%%%%%%%%%%%%%%%%%%%%%%%%%%%%%%%%%%%%%%%%%%%%%
%Equation: Perturbation B^z_eff leading term
\begin{equation}
 B^z_{k,eff}=\frac{4 \eta} {\Omega_0} \mathop{ \sum_{ i \neq k}} V_{ki}^{zx}  B^x
\label{dq:Bz_leading_term}
\end{equation}
%%%%%%%%%%%%%%%%%%%%%%%%%%%%%%%%%%%%%%%%%%%%%%%%%%%%%%%%%%%%%%%%%%%%%%%%%%%%%
\par
The expressions in Eqs.(\ref{eq:deltaE_leading_term}),(\ref{dq:Bz_leading_term}) are a result of an approximation which takes into account only the first excited electronic state $\Gamma_2$. Taking into account all excited electronic state we find a correction which amounts to multiplying $\delta E$ by $1.47$ (see details in Ref.~\onlinecite{YoavThesis}). Further quantitative accuracy comes from the calculation of the term linear in $B^x$ in third order of the perturbative expansion.

%%%%%%%%%%%%%%%%%%%%%%%%%%%%%%%%%%%%%%%%%%%%%%%%%%%%%%%%%%%%%%%%%%%%%%%%%%%%%
%Equation: deltaE to third order
%
This term is given by
\begin{equation}
\delta E^{(3)}= 2 g_L \mu_B \mathop{ \sum_{k}}  \left[ \mathop{ \sum_{j\neq k}} \mathop{ \sum_{i\neq j}} \frac{10 \eta^2}{\Omega_0^2} V_{kj}^{zx} V_{ij}^{xx} B^x \right] <J_k^{z}>
\end{equation}
%%%%%%%%%%%%%%%%%%%%%%%%%%%%%%%%%%%%%%%%%%%%%%%%%%%%%%%%%%%%%%%%%%%%%%%%%%%%%

We thus arrive at a quite accurate prediction for the energy splitting at small transverse fields. Considering only Ho ion pairs as in the numerical calculation we get:
%%%%%%%%%%%%%%%%%%%%%%%%%%%%%%%%%%%%%%%%%%%%%%%%%%%%%%%%%%%%%%%%%%%%%%%%%%%%%
%Equation: Perturbation DeltaE 2nd+3rd order
\begin{equation}
\begin{aligned}
\delta E_{perturbative}&=\delta E^{(2)}+\delta E^{(3)}=\\
=4 g_L \mu_B & \left[ \frac{4 \eta \tilde{V}^{zx}_{pair}}{\Omega_0} \left( 1.47 + \frac{10 \eta \tilde{V}^{xx}_{pair}}{4 \Omega_0}    \right)  B^x \right] <J_k^{z}>\\
&=0.8135 B^x \text{K}
\end{aligned}
\label{Bxanalytical}
\end{equation}
%%%%%%%%%%%%%%%%%%%%%%%%%%%%%%%%%%%%%%%%%%%%%%%%%%%%%%%%%%%%%%%%%%%%%%%%%%%%%
where the numerical value is given for n.n. pairs.

We note that terms which are third power in $B^x$ come from orders $4$ and higher of the perturbative expansion, and are therefore small. Our result in Eq.(\ref{Bxanalytical}) is in good agreement with our numerical calculations, which are well fit by the function
${\Delta E_{numerical}=[0.81 B^x -0.01 (B^x)^3] }$K.

%%%%%%%%%%%%%%%%%%%%%%%%%%%%%%%%%%%%%%%%%%%%%%%%%%%%%%%%%%%%%%%
%%%%%%%%%%%%%%%%%%%%%%%%%%%%%%%%%%%%%%%%%%%%%%%%%%%%%%%%%%%%%%%
%%%%%%%%%%%%%%%%%%%%%%%%%%%%%%%%%%%%%%%%%%%%%%%%%%%%%%%%%%%%%%%
\section{Numerics}
\label{numerics}
To diagonalize the 18496 X 18496 two ions Hamiltonian (17 electronic states times 8 nuclear states for each ion, squared for the two ions) for various pairs we use the Arnoldi method \cite{Arnoldi_complexity} (closely related to the Lanczos method \cite{Lanczos_Laflorencie}) . We do this in the transverse field range $0<B^x<2$T in increments of $10$mT. This iterative method is efficient in both computation time and storage space. Only a small fraction of the eigenstates is sought (we find the 576 lowest energy eigenstates which at zero transverse field correspond to states for which both ions are at one of the three lowest energy single ion electronic states ${|\uparrow>, |\downarrow>, |\Gamma_2>}$), and only the non-zero components of the sparse Hamiltonian matrix are stored (instead of the full matrix which takes up around 5GB of RAM and requires high-end hardware).
To calculate the effective field for a pair of ions at a given relative position we are interested in the energy difference between the states that at zero field are are defined as ${| \uparrow -\frac{7}{2} \uparrow -\frac{7}{2}>}$ and ${| \downarrow \frac{7}{2} \downarrow \frac{7}{2}>}$ (blue circles in Fig. \ref{fig:numerical_energy_levels_nn}). Values for the energy difference of other electronuclear pairs such as ${| \uparrow -\frac{7}{2} \uparrow -\frac{5}{2}>}$ and ${| \downarrow \frac{7}{2} \downarrow \frac{5}{2}>}$ are only slightly different.

As the field increases the states mix, yet, their Ising character given by their $<J^z>=\pm const$ value is well satisfied. Competition between the HF interaction and the effective longitudinal field result in level crossings of two spin states belonging to various nuclear states, see Fig. \ref{fig:numerical_energy_levels_nn}. For the calculation of the effective longitudinal field we follow the levels diabatically through the level crossings. Both the Ising character of the states and the diabatic tracking are well defined up to $B_x \approx 1.2$T. This value of the applied transverse field thus constitutes an upper limit to our calculations of the effective longitudinal field.

%\bibliography{RF}

\end{document}